\documentclass[12,final]{elsarticle}

\usepackage{graphicx,amsmath,amssymb,amsfonts}   % Include figure files
\usepackage{dcolumn}    % Align table columns on decimal point
\usepackage{bm}         % bold math
\usepackage{color}
\usepackage{times,setspace,multirow}

\newcommand{\ie}{\emph{i.e.\@} }
\newcommand{\etal}{\emph{et al.\@}}

\newcolumntype{d}[1]{D{.}{.}{#1}}
\begin{document}
\title{Energy spectra of primary knock-on atoms under neutron irradiation}

\author{M.R. Gilbert\(^1\), J. Marian\(^2\), J.-Ch. Sublet\(^1\)\\
\(^1\)Culham Centre of Fusion Energy, Culham Science Centre,\\Abingdon, OX14 3DB, UK\\
\(^2\)Department of Materials Science and Engineering, University of California Los Angeles, Los Angeles, CA 90095, USA}

\begin{abstract}
Materials subjected to neutron irradiation will suffer from a build-up of damage caused by the displacement cascades initiated by nuclear reactions. Previously, the main ``measure'' of this damage accumulation has been through the displacements per atom (dpa) index. There are known limitations associated with the dpa quantity and its domain of application and therefore this paper describes a more rigorous methodology to calculate the primary atomic recoil events (often called the primary knock-on atoms or PKAs) that lead to cascade damage events as a function of energy and recoiling species for any simulated or measured neutron irradiation scenario. Via examples of fusion relevant materials, it is shown that the PKA spectra can be complex, involving many different recoiling species, potentially differing in both proton and neutron number from the original target nuclei, including high energy recoils of light emitted particles such as \(\alpha\)-particles and protons. The variations in PKA spectra as a function of time, neutron field, and material are explored. Example PKA spectra are applied to radiation damage quantification using the binary collision approximation and stochastic cluster dynamics, and the results from these different approaches are discussed and compared.

\end{abstract}


\maketitle
\section{Introduction}
Understanding through modelling of the damage accumulated in materials irradiated by neutrons remains a primary goal for computational simulation of materials for advanced nuclear energy systems. Several different approaches exist for predicting the formation, evolution and behaviour of this damage, including: computationally demanding molecular dynamics simulations of damage cascades with full atomic interactions; rate theory models, where defects are described as objects with defined behaviour; and kinetic Monte-Carlo (kMC) simulations. However, without exception, all of these techniques need, as input at some level, information about the initial primary disruptions, in the form of the type, energy and spatial distribution of the primary knock-on atoms (PKAs). In particular, this kind of data goes far beyond the limited information provided by the traditional and ubiquitous defect index of initial damage formation known as ``displacements per atom'' (dpa). The so-called Norgett, Robinson and Torrens NRT-dpa~\cite{norgettetal1975} reduces the predicted irradiation environment, perhaps obtained from a Monte-Carlo neutron transport simulation on a reactor geometry, to a single number that converts the energy deposited into a material by the irradiation into an estimate of the number of atomic displacements that could be generated. It is obvious that such a measure should not be used as a basis for comparing irradiation behaviour in materials~\cite{odettedoiron1976}. A more complete picture of radiation damage evolution, that may be afforded by modern computational techniques, requires the spatial and energy distribution of all the initial displacement events, including both emitted and residual nuclei from nuclear interactions.

In modern nuclear data evaluations, for a single target species such as the major \(^{56}\)Fe isotope in Fe, there are many nuclear reaction channels that produce recoiling species, including elastic, inelastic, and nonelastic nuclear reactions. In nuclear fusion in particular, the generally higher energy of the incident neutrons, when compared to fission, leads to many more channels becoming relevant, which in turn produces a more complex distribution of PKAs in both energy and type. These PKAs lead to cascades of atomic displacements, which can subsequently evolve and collapse to produce extended defects such as dislocation loops and voids.

In this paper a modern computational methodology is described to produce, for a given target species or distribution of targets, the instantaneous picture of PKA rates as a function of energy (``PKA spectra''). This includes a newly written code, called \texttt{SPECTRA-PKA}, that takes as input nuclear recoil cross-section matrices and combines (collapses) these with an incident neutron spectrum to produce the PKA distributions.
PKA spectra for all the primary (heavy) nuclear reaction products, also known as the residuals, which may be a different elemental species than the target nuclide, and any secondary (light) emitted products are considered. Results for selected fusion-relevant materials under various neutron irradiation fields are presented. In the final section outputs from the above are used as input to two applications for radiation damage characterisation and quantification: the binary collision approximation (BCA) calculations and stochastic cluster dynamics (SCD) simulations.

\section{Methodology and Results}\label{methodology}

The nuclear data processing code NJOY~\cite{njoy1983}, and in particular the GROUPR module within it, can calculate group-to-group recoil cross-section matrices due to many types of nuclear reactions. Using neutron-interaction data for a given target nuclide \(x\), such as \(^{56}\)Fe, NJOY-12~\cite{njoy2012} has been used in the present work to provide matrices \(M^{x\rightarrow y}\)  for every \(x\rightarrow y\) reaction channel. Group-to-group cross-section matrices for neutron scattering, light charged particles, as well as recoils of the residual nuclei can be generated from modern nuclear data libraries, following the ENDF~\cite{endf605} data format. The evaluations include both energy transfer and angular recoil distributions for all energetically possible nuclear reactions on a wide range of target nuclides, which NJOY reads and processes into a pre-defined group structure.

Two-body elastic and discrete inelastic neutron scattering, charge-particles elastic scattering, continuum scattering and fission can be treated in different ways depending on the completeness and accuracy of the content of the original evaluation. Previously, simple theoretical models had to be employed to plug the gaps in earlier nuclear data libraries, but this is becoming less necessary with the latest libraries~\cite{tendl2014}, which use detailed and well-validated theoretical models to derive data where no experimental information exists. The resulting \(m^{x\rightarrow y}_{ij}\) elements of each \(M^{x\rightarrow y}\) matrix are the cross-sections in barns for a recoil in energy group \(i\) produced by an incident neutron in energy group \(j\).

A fine 709-group~\cite{subletetal2012} spectrum has been used here for both the incident and recoil energy distributions, and all the results were calculated using input data from the TENDL-2014~\cite{tendl2014} nuclear library, which was selected in part because it contains the extended database of nuclides (compared to other libraries) required for complex material compositions (see Section~\ref{realmaterials}), whilst providing the complete and often intricate energy-angle distributions necessary for this type of simulation.

Note that for the present work the detailed angular distributions of the recoiling nuclei and scattered particles, which are also calculated by NJOY, are not explicitly retained, although the angular dependence is implicitly considered in as far as it impacts on the recoil energy distribution. It is assumed that any neutron-incident spectrum used in conjunction with the recoil cross-section matrices will be average fluxes over relatively large (on the atomic scale) volumes. In this case both the lack of directional information in the neutron field and the absence of structural information about the irradiated materials, including the distribution of grain orientations, means that including direction information in a PKA spectrum is not relevant. An isotropic recoil distribution is considered a valid approximation for this level of simulation. If, in the future, it is possible to accurately define the orientation of the atomic lattice relative to the neutron-irradiation source, perhaps in a well-qualified fusion reactor first wall (FW), then it may be necessary to revisit this approximation and consider the angular dependence of recoils.

\begin{figure}[h]
%\frame
{\includegraphics[width=1.0\textwidth]
{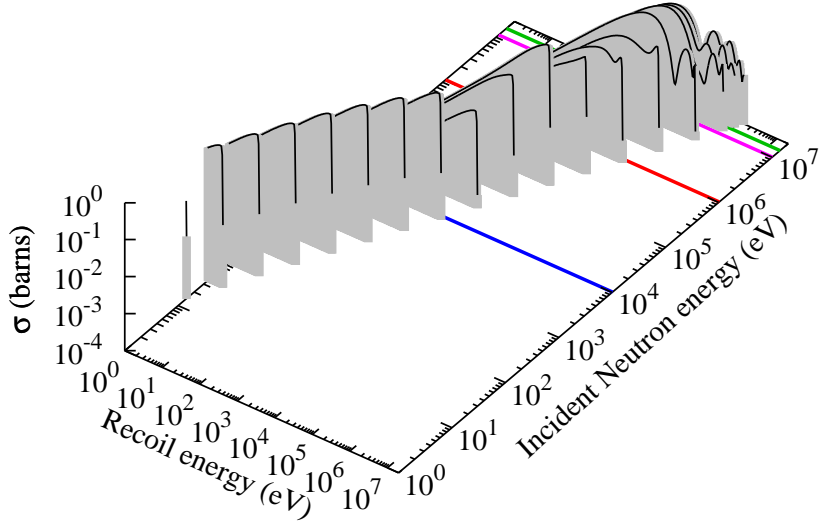}}
\caption{\label{Fe56_scatter_matrix}The recoil cross-section matrix for elastic scattering of neutrons on \(^{56}\)Fe. The matrix is plotted as a separate recoil-energy vs. cross-section distribution for each incident energy group (plotted at the midpoint of the group). Note that only a subset of the possible 709 incident-energy-group distributions are shown. The coloured lines on the base xy-plane indicate the four incident energies considered in Fig.~\ref{Fe56_matrix_2d} to display snapshots of the recoil cross-sections for multiple reaction channels. (For interpretation of the references to colour in this figure caption, the reader is referred
to the web version of this article.)}
\end{figure}

\begin{figure}[h]
%\frame
{\includegraphics[width=1.0\textwidth]
{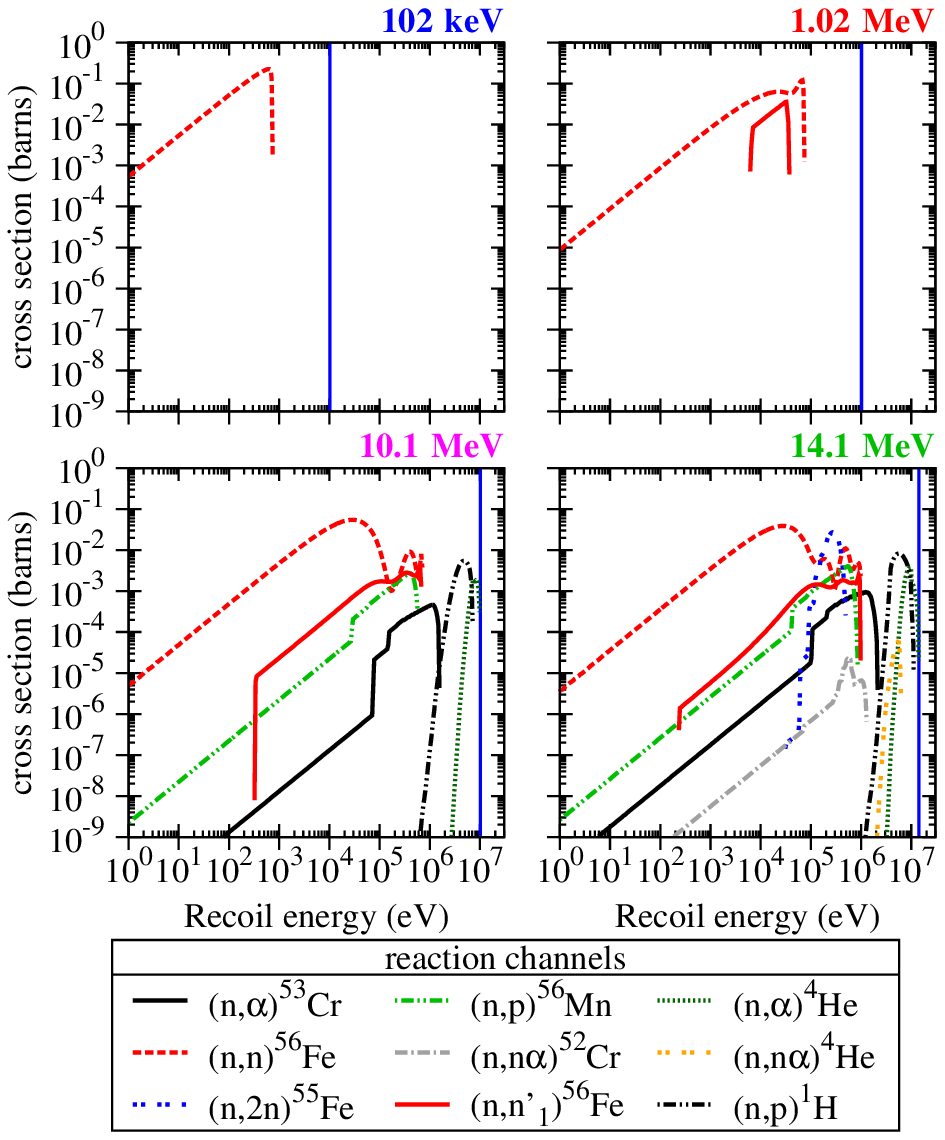}}
\caption{\label{Fe56_matrix_2d}Incident energy-group snapshots of the group-to-group recoil matrices generated using NJOY for reactions on \(^{56}\)Fe. The four plots show the cross-sections as a function of recoil energy for different incident neutron energies groups, whose midpoint energies are indicated by the solid vertical line (blue)  and the label in the top right-hand corner of in each plot. The colouring of the text labels is equivalent to that used to indicate these same incident energies on Fig.~\ref{Fe56_scatter_matrix}. Note that here, and in the other figures in this paper, y-values are plotted against the midpoints of the energy group to which they correspond. (For interpretation of the references to colour in this figure caption, the reader is referred
to the web version of this article.)}
\end{figure}

Fig.~\ref{Fe56_scatter_matrix} is a 3D plot of the elastic scattering \((n,n)\) recoil cross-section matrix for \(^{56}\)Fe, while
Fig.~\ref{Fe56_matrix_2d} shows 2D snapshots at several incident energy groups for the main reaction channels of the same nuclide, including elastic scattering. For each snapshot in Fig.~\ref{Fe56_matrix_2d} at incident energy group \(j\) (indicated by a vertical line in the group mid-point in each graph and by the straight lines on the xy-plane of Fig.~\ref{Fe56_scatter_matrix}) the curves effectively represent the \(j\)th column vectors of the \(M^{x\rightarrow y}\) matrices. The figure shows that there is significant variation as a function of neutron energy for this \(^{56}\)Fe case, in terms of both the reaction channels represented and the cross-sections versus recoil-energy distributions of the different channels. At low neutron energies around 102~keV, only a single reaction channel is open -- that of the simple elastic scattering, where the same neutron is emitted as was incident (hence the \((n,n)\) representation in the figure). As the neutron energy increases, other channels become important, from inelastic reactions (here only the inelastic channel to the ground state \((n,n'_1)\) is shown) and then more complex non-elastic reactions, where a variety of different particles are emitted.

Fig.~\ref{Fe56_matrix_2d} also demonstrates that NJOY produces the cross-sections for the secondary light particle emissions associated with some of the reaction channels. For example, in addition to the recoils of \(^{53}\)Cr from the \((n,\alpha)\) reaction, there are also cross-section curves for the recoil of the \(^{4}\)He \(\alpha\)-particles from the same reaction. Note that in Fig.~\ref{Fe56_matrix_2d} there are no curves associated with the main \((n,\gamma)\) neutron-capture channel because the version of NJOY used (2012-032), does not directly output  the matrices for the recoils of this channel. The method by which the recoils from this important channel are calculated in the present  work is discussed later.

%\begin{figure}[h]
%%\frame
%{\includegraphics[width=1.0\textwidth]
%{W-184_recoil_vectors.eps}}
%\caption{Incident energy-group snapshots of the group-to-group recoil matrices generated using NJOY for reactions on %\(^{184}W\). The four plots show the cross-sections as a function of recoil energy for different incident neutron energies %groups, whose midpoint is given in the top right of each plot. This midpoint energy is also indicated by the solid vertical %line (blue) in each plot. Note that here, and in the other figures in this paper, y-values are plotted against the midpoints %of the energy group to which they correspond.}
%\end{figure}

%\clearpage

To calculate PKA spectra, these recoil cross-section matrices must now be folded with a neutron flux versus energy spectrum for the irradiation environment of interest. The code SPECTER and its sister code SPECOMP have previously been developed at Argonne National Laboratory by Greenwood and Smitther~\cite{specter} to perform this folding or collapsing for both pure elements and compounds, respectively. However, both codes are relatively old, having been written in the 1980s, and are limited in their ability to handle modern nuclear data and complex material compositions. For this reason a new code has been developed, hereafter called \texttt{SPECTRA-PKA}, which can take group-wise recoil matrices generated directly from NJOY (with a slight modification to the GROUPR routine therein to separate the recoil matrices from other NJOY data) and collapse them with a user-defined neutron-irradiation spectrum. In particular, the new code can take both recoil matrices and neutron spectrum in any user-defined energy group structure, a feature not available in SPECTER/SPECOMP, allowing the use of the fine 709-group structure mentioned previously. The new code also performs interpolations and averaging, as necessary, between different spectrum energy-group formats -- for example if the neutron irradiation spectrum has been computed with a different format to that used for the recoil cross-section matrices -- although we recommend using the same high-resolution 709-group (see~\cite{subletetal2012} for details) structure to calculate irradiation spectra, particularly when using statistical Monte-Carlo methods.

A further advance over the previous code is the ability, in \texttt{SPECTRA-PKA}, to consider multiple reaction channels on the same target. Typically, NJOY will process the entire data file for a given target nuclide and output recoil matrices for every reaction channel in the same file. Our new code will read this file and process each channel one-by-one, outputting the PKA matrix that results from each folding. For a given \(x\rightarrow y\) reaction channel, the PKA spectrum \(R^{x\rightarrow y}\) is computed via:
\begin{equation}
 R^{x\rightarrow y}(E)\equiv\{r^{x\rightarrow y}_i\}=\left\{\sum_jm^{x\rightarrow y}_{ij}\phi_j\right\},
\end{equation}
where \(r^{x\rightarrow y}_i\) is the PKA rate in recoil energy-group \(i\), computed by folding the \(i\)th row of \(M^{x\rightarrow y}\) in cross-section units of barns (\(1\times10^{-24}\)~cm\(^{2}\)) with the \(\phi_j\) flux values of the incident neutron spectrum in units of neutrons~cm\(^{-2}\)~s\(^{-1}\).

\begin{figure}[h]
%\frame
{\includegraphics[height=0.7\textwidth]
{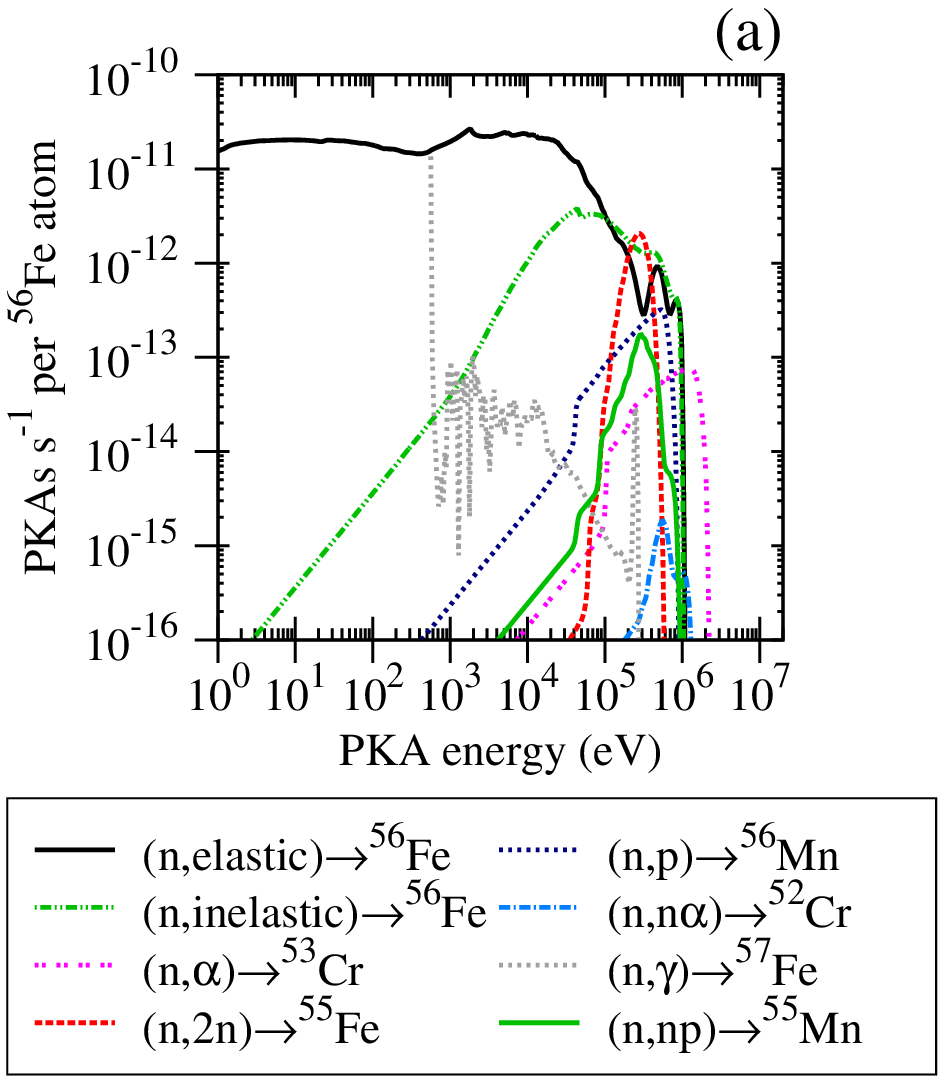}}
%\frame
{\includegraphics[height=0.7\textwidth]
{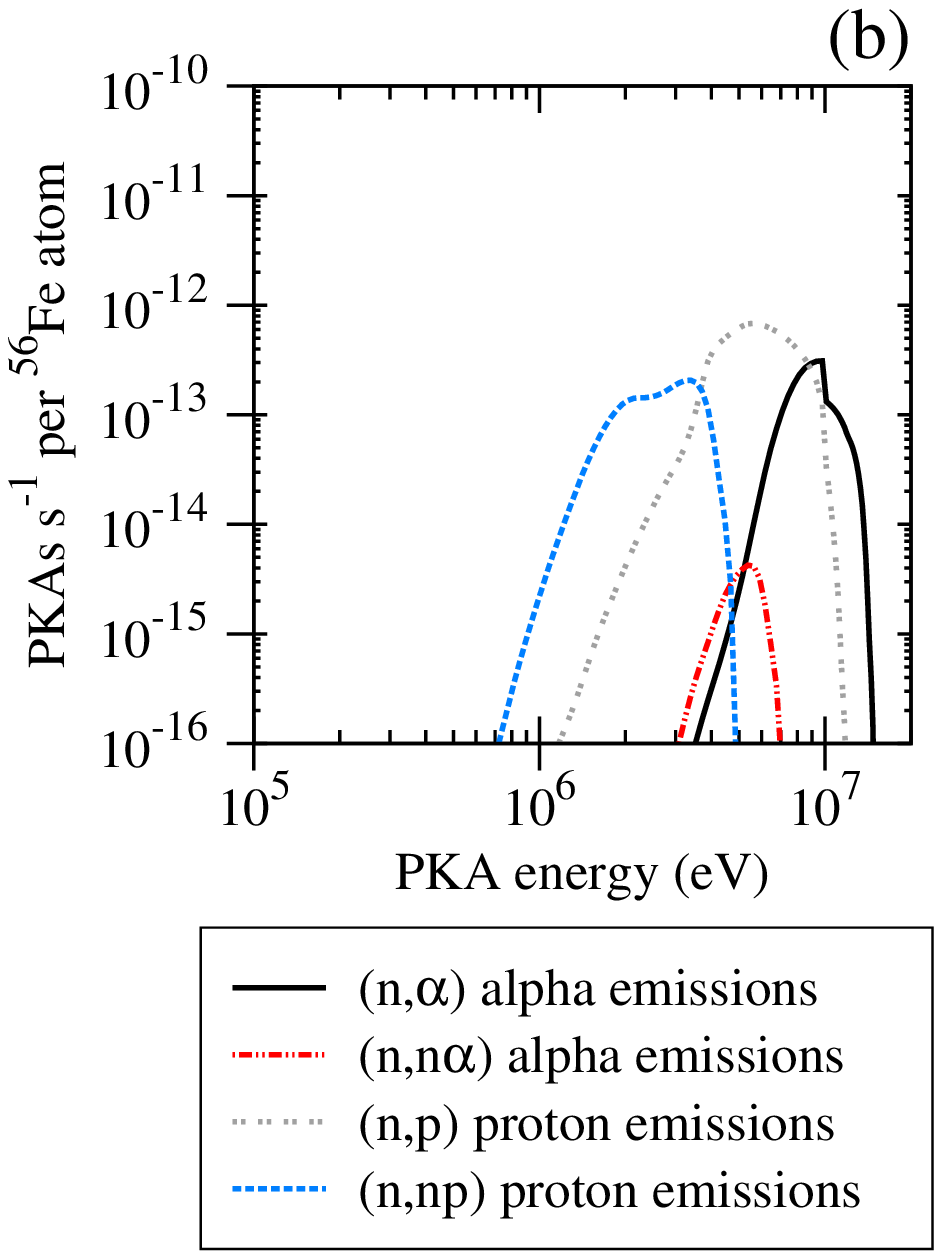}}
\caption{\label{Fe56_channels}The collapsed PKA spectra for the dominant reaction channels in \(^{56}\)Fe under DEMO hcpb FW conditions. (a) the (heavy) residual recoils, and (b) the secondary light particle emissions. Note the change in energy scale in (b) reflecting the fact that the light particles are emitted with a higher proportion of the recoil energy than the associated heavy residuals produced via the same reaction channels.}
\end{figure}

\begin{figure}
%\frame
{\includegraphics[width=1.0\textwidth,clip=true,trim=0cm 0cm 0cm
0cm]{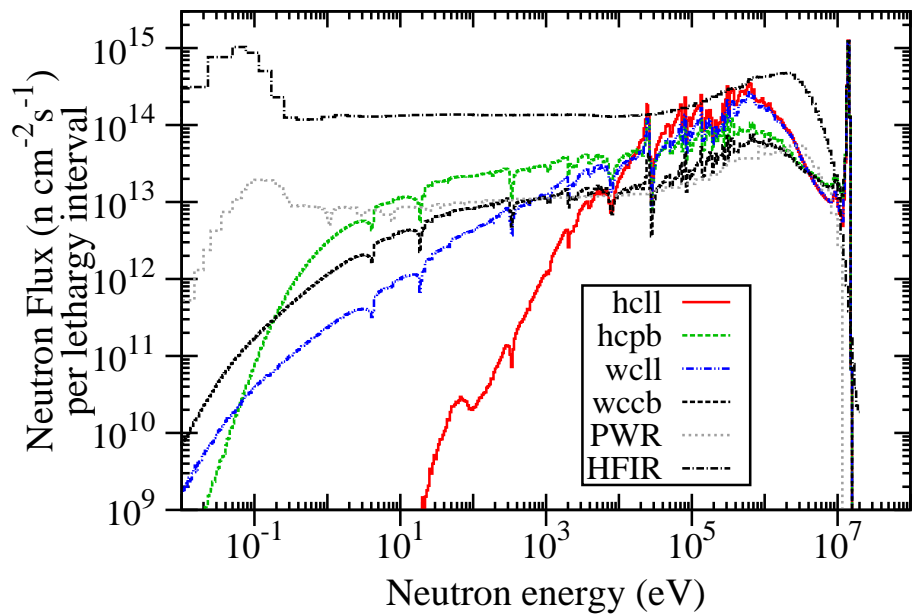}}
\caption{\label{spectra}Neutron irradiation spectra for the equatorial (outboard) FW of several demonstration fusion reactor (DEMO) concepts~\cite{gilbertetalfst2014}. Also shown, for comparison, is the fuel-assembly averaged spectra for a typical fission reactor -- in this case for a pressurised water reactor or PWR -- and also the spectrum predicted for the HFIR experimental fission reactor~\cite{greenwood1994}. The neutron flux is plotted per lethargy interval, which is equal to the natural logarithm of the upper divided by lower energy bounds of each of energy group used to bin the spectrum.}\end{figure}

Fig.~\ref{Fe56_channels}, shows the PKA spectra computed for the main reaction channels on \(^{56}\)Fe under a  neutron-irradiation flux-spectrum predicted for the equatorial first wall (FW) of a typical conceptual design for a demonstration fusion power plant ``DEMO''. This spectrum, shown in Fig.~\ref{spectra}, was computed using the MCNP~\cite{mcnp1,mcnp2} Monte-Carlo neutron transport code for a 2013 DEMO design with helium-cooling and a tritium-breeding blanket made up of a bed of Li+Be pebbles, designated as hcpb in the figure for helium-cooled pebble-bed -- see \cite{gilbertetalfst2014} for further details of the model and calculations. Fig.~\ref{Fe56_channels}a shows the PKA rates in units of PKAs per second per target \(^{56}\)Fe atom for the main heavy nuclide recoil channels, while  Fig.~\ref{Fe56_channels}b shows the equivalent light particle (gas) emission PKA spectra, where they exist.

\clearpage


Fig.~\ref{Fe56_channels}a shows that for much of the PKA energy range (on a logarithmic scale), the simple elastic scattering or \(^{56}\)Fe atoms dominates, which is the typical result for most nuclides, with PKA rates of the order of \(2\times10^{-11}\)~PKAs~s\(^{-1}\) per target atom at all energies below a few tenths of a keV. Only at energies above around 100~keV do other reaction channels produce statistically significant numbers of PKAs. Most notably the inelastic scattering channel dominates the elastic scattering above \(\sim\)100~keV, while  \(^{53}\)Cr PKAs from the \((n,\alpha)\) reaction are the only significant heavy recoils above 1~MeV. Note in this latter case that the recoil nuclei is of a different chemical nature to the target nuclei because transmutation has occurred.

For the light particle emissions (Fig.~\ref{Fe56_channels}b) note that these are predominantly at energies above 1~MeV, reflecting the fact that, via conservation of momentum, they receive a higher proportion of the energy from the 14~MeV neutrons that dominate the irradiation spectrum (see Fig.~\ref{spectra}) than the heavy recoils generated in the same nuclear reactions. The PKA rates for these light particles are also relatively low compared to the rates observed for the heavy-particle scattering, but they are in agreement with the rates for the corresponding heavy recoils for the same reaction channels.

Fig.~\ref{Fe56_channels}a  also includes the PKA spectrum associated with the recoil of \(^{57}\)Fe, which is produced by neutron-capture \((n,\gamma)\) followed by photon (\(\gamma\)) emission. As mentioned earlier, the recoil matrix for this channel is not an automatic output from GROUPR. However, GROUPR produces the \((n,\gamma)\) reaction cross-section vector from the same nuclear data and in the correct group format. There is no energetic cost associated with this type of reaction, and furthermore if we assume that no energy is lost through excitation of the residual, then a simple momentum conservation argument gives the energy range of the target atom once it has captured a neutron from a specific energy group. The resulting compound nucleus will still be in an excited state due to extra mass and must decay down the energy levels by emitting photons (\(\gamma\)-rays), whose total energy can be found by comparing the mass of the final ground state of the daughter atom, in the fig.~\ref{Fe56_channels} case \(^{57}\)Fe, with the combined mass of target (\(^{56}\)Fe) and neutron (the ``compound nucleus''). Then, as a conservative (in the sense that the highest possible recoil energy is attained) approximation, it is assumed that all of this energy, via Newton's Third law, results in additional momentum. An alternative to this maximum energy transfer approximation (see~\cite[Chap. 1]{was2007}), is to take the average and use half of this energy (also suggested in~\cite{was2007}). Continuing with the conservative approach, for \(^{56}\)Fe(\(n,\gamma\))\(^{57}\)Fe this results in approximately 551~eV extra for every recoil energy, which, once collapsed with a neutron spectrum, results in a PKA spectrum that has no recoils below this.

In the analysis and calculations presented above and hereafter we do not consider any additional recoils that might be produced through radioactive decay. Some of the residual nuclides, even in their ground-state, will be unstable (e.g. the \(^{55}\)Fe produced via \((n,2n)\) reactions on \(^{56}\)Fe)  and will decay via the emission of further energetic species, such as \(\beta\) particles. This will, of course, via momentum conservation, cause the residual decayed nuclei to recoil. Furthermore, depending on the half-life, the decay may happen on the same timescale of the primary recoil from the nuclear reaction and would thus increase the overall recoil energy. Alternatively, the decay may happen long after the radioactive species has come to rest and thus would initiate a new, independent recoil. While it is expected that the recoil energies associated with these decays will generally be lower in energy than that created by the nuclear interaction and significantly less frequent, they should still be included for completeness and we are investigating how this can be done most effectively. Note, that in the earlier code, SPECTER~\cite{specter}, the (\(n,\gamma\)) reaction channel was treated with subsequent \(\beta\) decay of the residual included in the kinematics, and this is an important channel in the nuclear fission industry, but for more general purposes, including fusion, we would need to consider all decay channels.

\subsection{Calculations for real materials}\label{realmaterials}

For a real material, even a single element with multiple isotopes, the results for a single target nuclide are not sufficient. Thus, \texttt{SPECTRA-PKA} has been written to process multiple sets of recoil cross-section matrices and perform various summing to produce PKA spectra in the most appropriate format for damage modelling. For each target species, each with its own separate input file of recoil cross-section matrices, the code folds the matrix for every channel with the neutron flux-spectrum, but then weights the resulting spectrum of PKA rates by a user-defined atomic fraction of the current target in the material composition. Subsequently, when the PKA rates for all targets have been processed in this way, the spectra are summed according to whether they are for the same PKA daughter. So, for example, in an irradiation of pure W, where \(^{182}\)W, \(^{183}\)W and \(^{184}\)W all form part of the natural composition, the total PKA spectrum for \(^{183}\)W, would include contributions from scattering (both elastic and inelastic) on \(^{183}\)W itself, but also from the neutron-capture (\(n,\gamma\)) reaction on \(^{182}\)W, and the (\(n,2n\)) reaction channel on \(^{184}\)W. Numerically, the total PKA spectrum for a given recoil species \(y\) is:

\begin{equation}
R^{y}(E)=\sum_{i,j}w_iR^{x_i\overset{c_j}{\longrightarrow} y}(E),
\end{equation}
where \(w_i\) is the atomic fraction of target \(x_i\) in the material of interest, and the contribution to the \(y\) PKAs from \(x_i\) is the sum over all possible reaction channels \(c_j\) that produce \(y\) from \(x_i\).

\begin{figure}
%\frame
{\includegraphics[width=0.5\textwidth]
{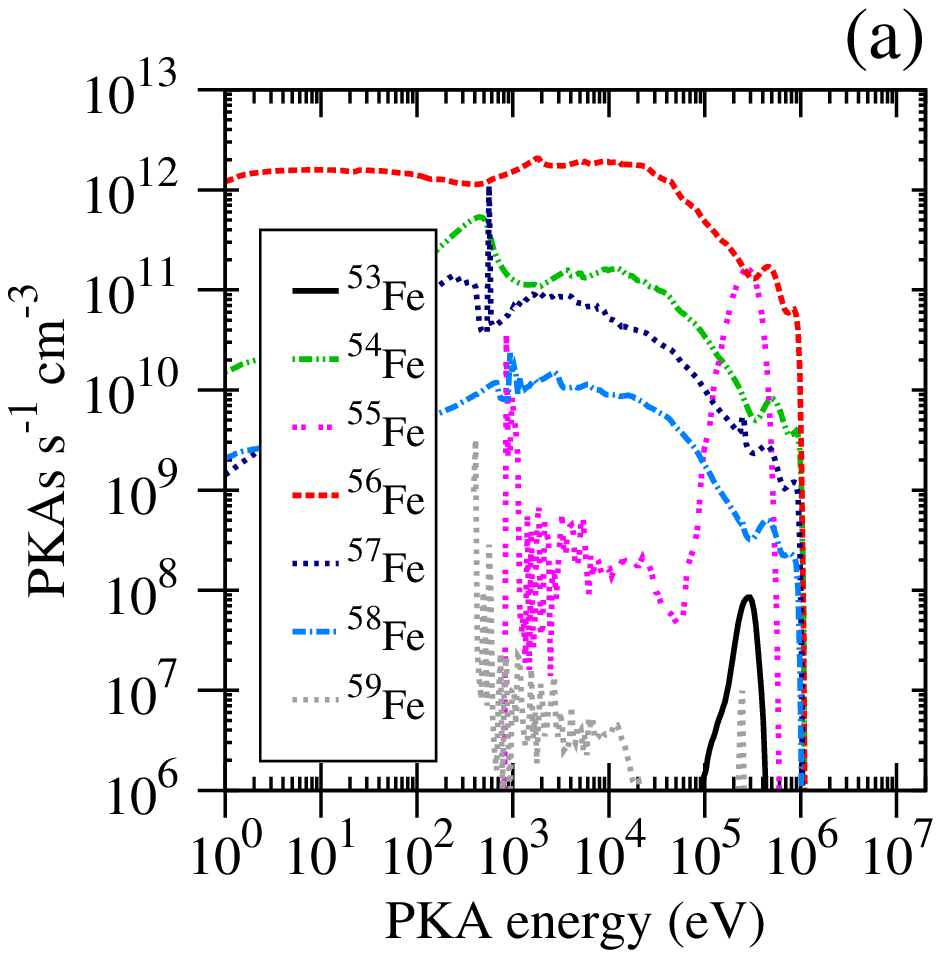}}
%\frame
{\includegraphics[width=0.5\textwidth]
{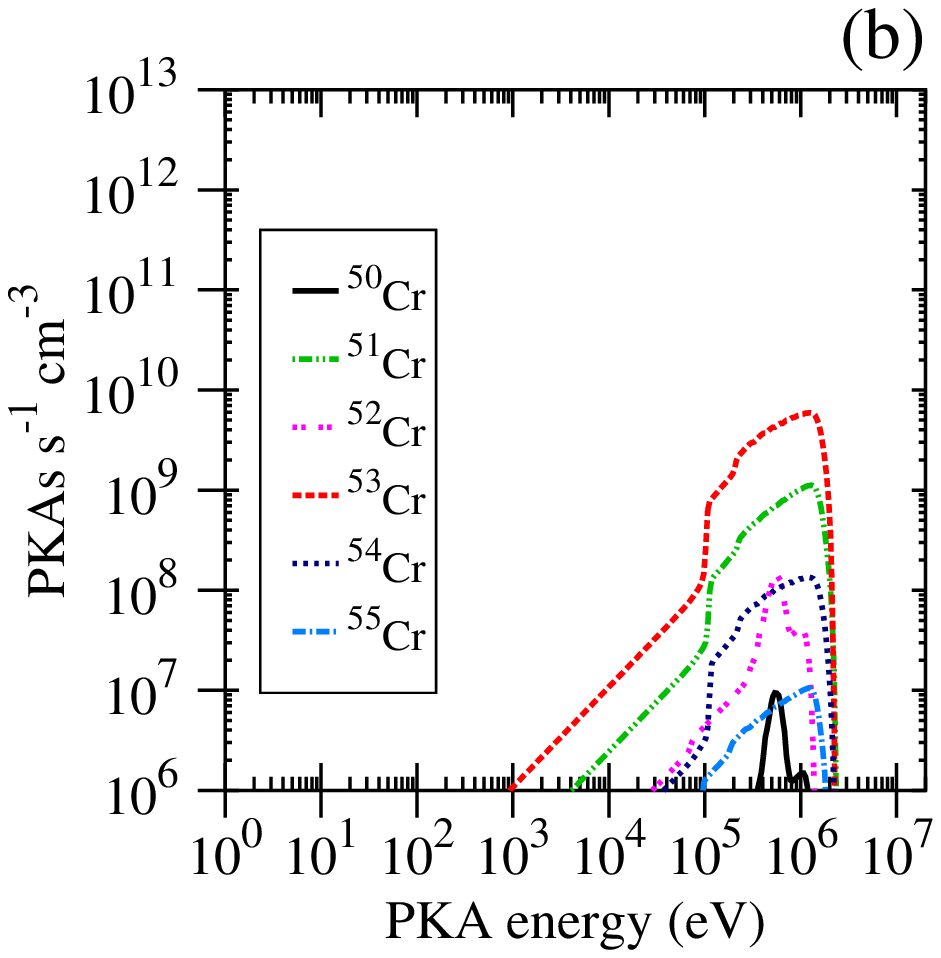}}
%\frame
{\includegraphics[width=0.5\textwidth]
{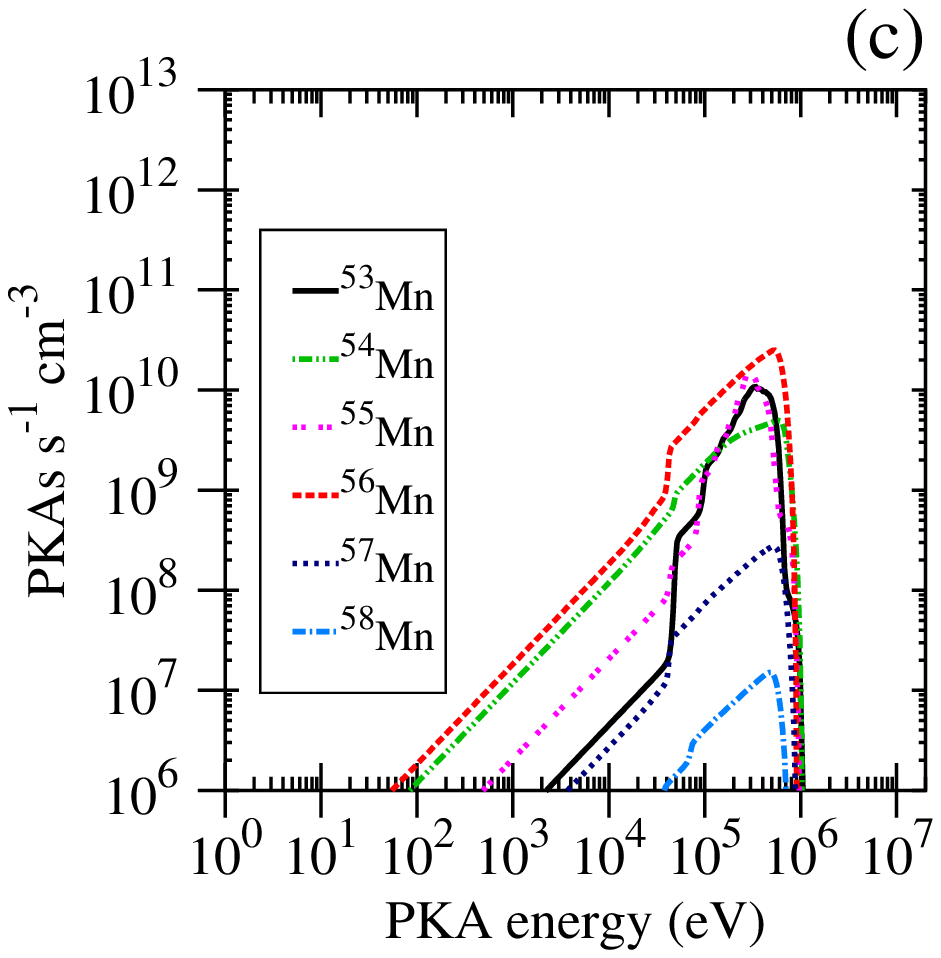}}
%\frame
{\includegraphics[width=0.5\textwidth]
{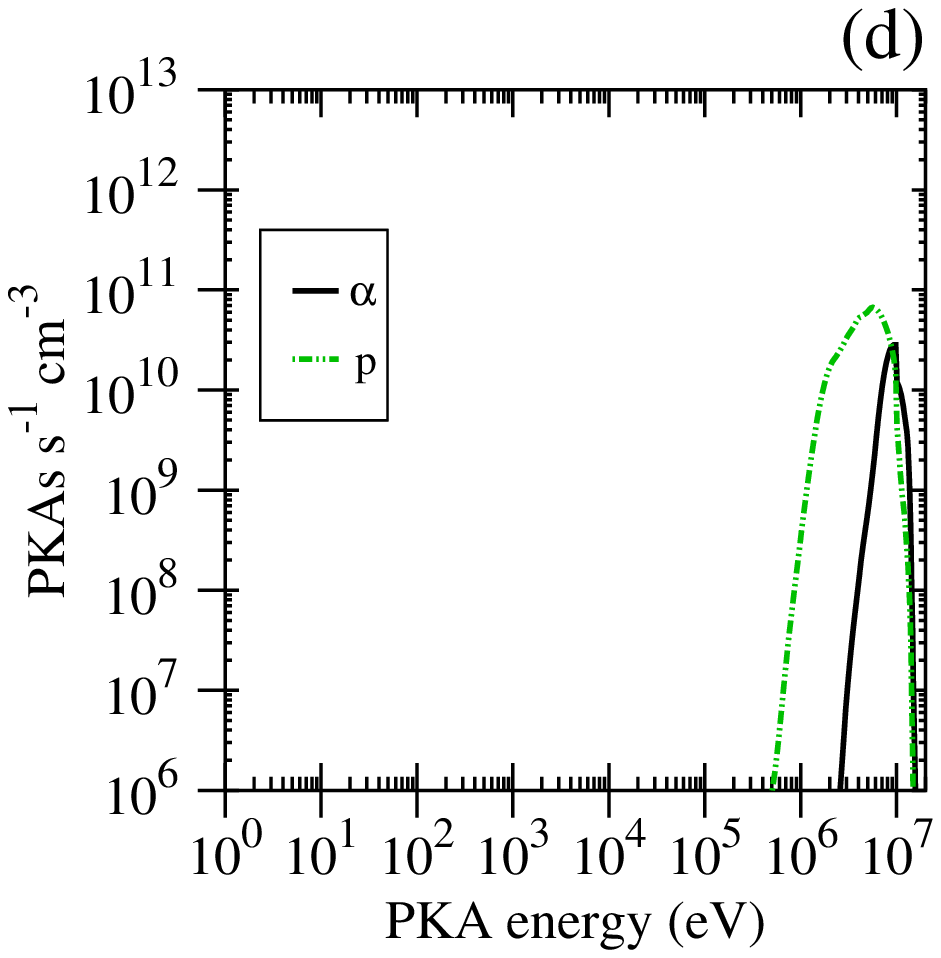}}
\caption{\label{Fe_isotopes}The set of PKA-rate spectra for pure Fe irradiated in the FW hcpb DEMO neutron-field (Fig.~\ref{spectra}). (a) shows the PKA spectra for the isotopes (both naturally occurring and those produced by nuclear reactions) of Fe, while (b) and (c) show the PKA spectra for the isotopes of Cr and Mn, respectively, generated by neutron capture followed by proton (for Mn) or \(\alpha\) (Cr) emission reactions on Fe. (d) shows the summed PKA spectra for the light particles themselves. The conversion from the original ``per target isotope'' units (as used in Fig.~\ref{Fe56_channels}) to PKAs~s\(^{-1}\)~cm\(^{-3}\) was performed using the density and mass of pure Fe given in Table~\ref{demohcpbfw}.
}
\end{figure}

For Fe, four isotopes make up its natural composition, with \(^{56}\)Fe the most common at 91.75 atomic \% abundance then \(^{54}\)Fe, \(^{57}\)Fe, and \(^{58}\)Fe at 5.85, 2.12, and 0.28 atomic \%, respectively. The resulting summed nuclide PKA spectra for the hcpb DEMO irradiation scenario are shown in Fig.~\ref{Fe_isotopes}, where the isotopes of Fe, Cr, and Mn are shown in three separate plots to allow easy identification, together with a fourth plot showing the summed PKA spectra for \(\alpha\) particles and protons. Note that while the output from the folding of recoil cross-section matrices with neutron flux spectra produces PKA rates with units of PKAs~s\(^{-1}\) per target atom, in this figure and subsequent ones in this paper the PKA rates are given for unit volumes of target material. This avoids the ambiguity associated with ``per target'' when there is a complex material composition, and is more useful for modelling where a defined volume of material is considered and hence the number of atoms is known, for example in a molecular dynamics simulation of a collision cascade. Note that this cm\(^{-3}\) measure is not related to the cm\(^{-2}\) employed when defining the neutron flux field that every target atom sees, but is a more convenient unit for the present results.

As expected, the highest PKA rates are calculated for the four naturally-occurring isotopes of Fe (Fig.~\ref{Fe_isotopes}a), with \(^{56}\)Fe dominating overall due to its high abundance in pure Fe combined with the large scattering cross-sections. PKAs associated with heavy Mn or Cr recoils (Figs.~\ref{Fe_isotopes}c and \ref{Fe_isotopes}b, respectively) are less frequent and generally only at higher energy due to the lower cross-section and threshold nature of the charged particle reactions (\((n,\alpha)\), \((n,p)\), etc), that produce them, although, for Cr in particular, the maximum PKA energy is higher than that of Fe PKAs. Finally, the recoils of light particles (Fig.~\ref{Fe_isotopes}d) are at even higher energies, in agreement with the earlier results for \(^{56}\)Fe, and with similar lower production rates as for Mn and Cr.

Two further merges of the results, which are also built into the \texttt{SPECTRA-PKA} processing code, are also relevant to materials modelling. Noting that simulating different isotopes of the same element is not normally considered in atomic simulations, it is more useful to consider the summed PKAs for each daughter element produced in the nuclear reaction, whether these are different from the original target elements or not. Furthermore, taking a more pessimistic view of simulations, unless the material under investigation is a genuine mixed alloy or compound, it is unlikely that the models used to describe atomic interactions will include parameters for every possible atom species generated under neutron irradiation, or even be able to consider the spontaneous transformation of one atom type into another. In this sense we are primarily limited to considering PKAs that are identical to the host lattice and so the code also outputs the total PKA spectrum for all recoils, by summing over all daughter elements. It is expected that this would be most useful in situations where the target material is a single element or simple alloy.

\begin{figure}
\includegraphics[width=0.5\textwidth]
{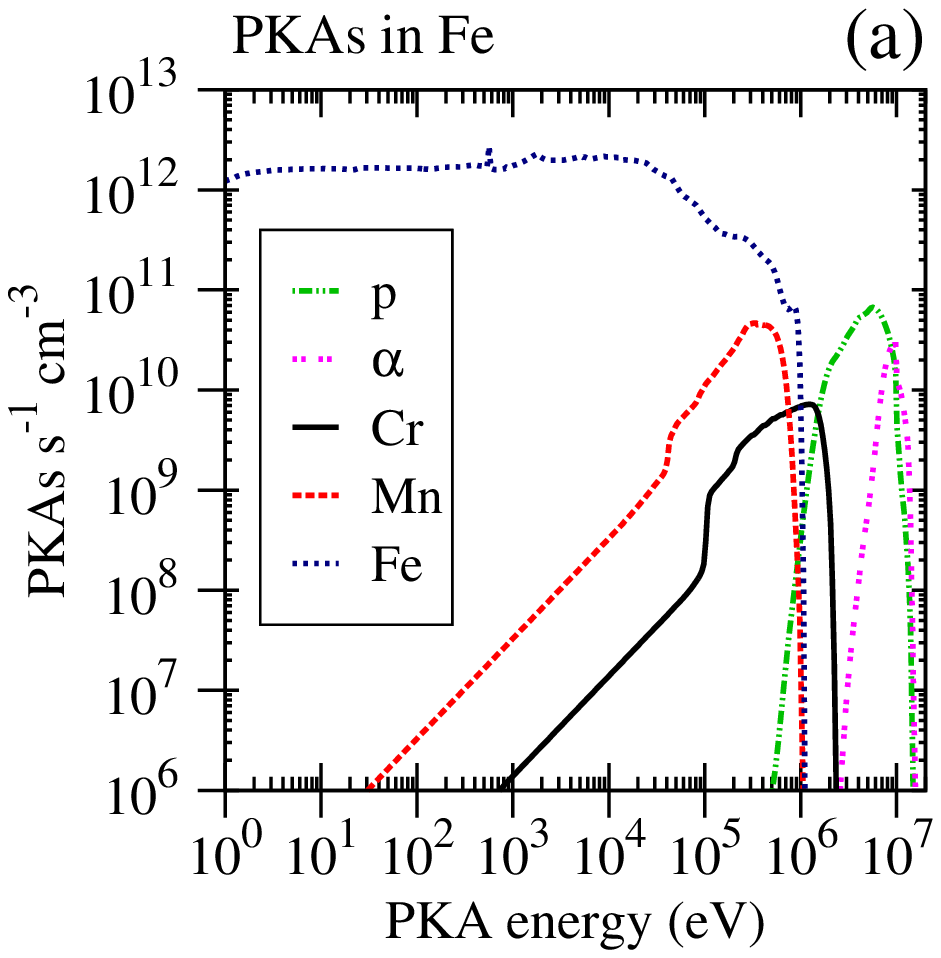}
\includegraphics[width=0.5\textwidth]
{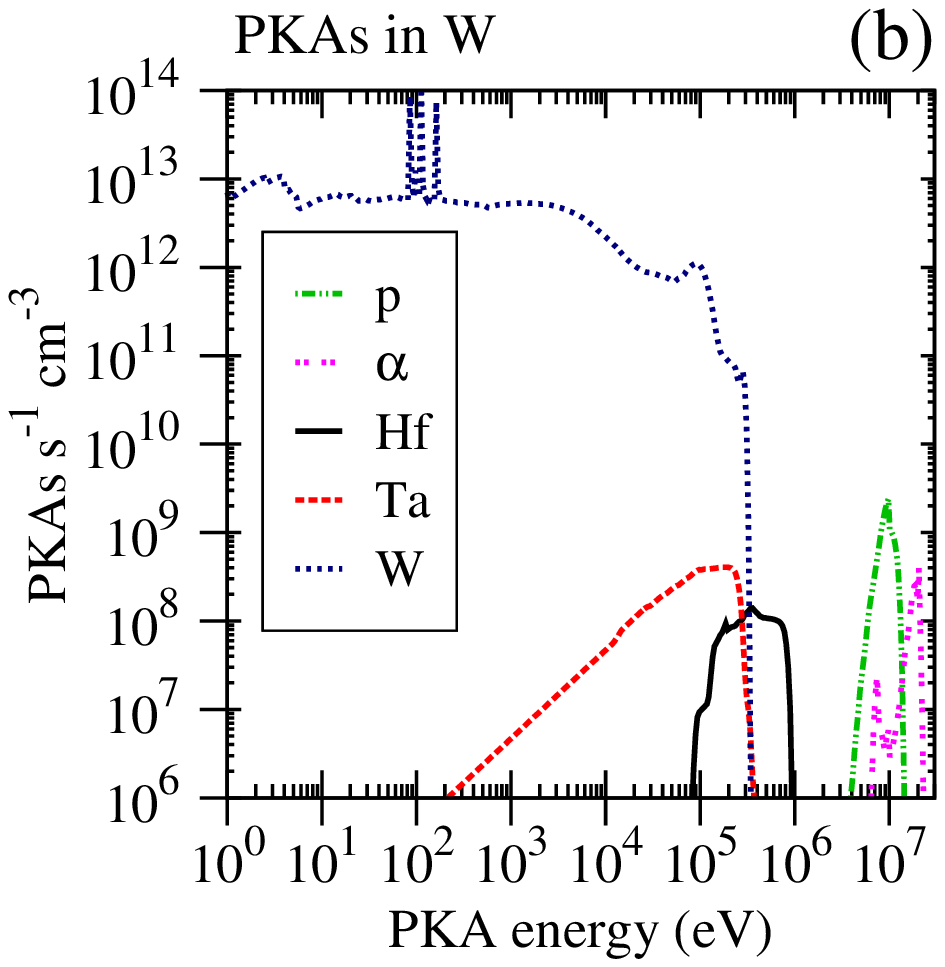}
\caption{\label{elemental} The set of summed elemental PKA spectra in pure (a) Fe and (b) W under the DEMO hcpb FW irradiation scenario (see Fig.~\ref{spectra}). The conversion to PKAs~s\(^{-1}\)~cm\(^{-3}\) was performed using the densities and masses of pure Fe and W given in Table~\ref{demohcpbfw}.}
\end{figure}

Fig.~\ref{elemental}a shows the summed elemental PKA spectra for pure Fe under the DEMO hcpb FW neutron spectrum, while Fig.~\ref{elemental}b has results for pure W, which has five natural isotopes (180, 182, 183, 184, and 186 at 0.12, 26.56, 14.31, 30.64, and 28.43 atomic \%, respectively), under the same conditions. Note that these PKA spectra, and those shown in previous figures, represent a snapshot at \(t=0\), i.e. before any neutron-induced transmutation has taken place. For Fe, where transmutation rates are relatively low under fusion neutron conditions, this is a reasonable representation of the picture throughout the lifetime of an irradiated sample. However, in a material such as W the transmutation rates are much higher and there could be a significant change in chemical composition during reactor and component lifetimes. In such circumstances, the change in chemical composition might lead, as a side-effect, to a discernable change in PKA spectra profiles, leading to a corresponding change in damage accumulation. This will be investigated in a later section, but note that a consequence of the \(t=0\) assumption in these figures, and elsewhere where no transmutation is considered, there are no PKA contributions from elements with a higher atomic number \(Z\) than the parent target atoms -- in this case Fe and W, respectively for Figs.~\ref{elemental}a and b. This is a natural consequence of the fact that there is no direct reaction channel with a neutron as the projectile that leads to an increase in \(Z\) -- some kind of decay process, such as \(\beta^{-}\) decay, is needed for that. As suggested earlier, the recoils of such decays may give a non-negligible contribution to the total PKA picture.

Fig.~\ref{elemental}, shows, for W in particular, a very startling result for the recoil energies of the \(\alpha\) particles, namely that the energies at which they can be emitted is significantly above the maximum neutron energy, in this case around 14~MeV. Indeed, the maximum energy of the emitted \(\alpha\) particles under this fusion spectrum is above 20~MeV, with the distribution peaking at around 18~MeV. The reason for the very high emission energy is that the Q-values for the \((n,\alpha)\) reaction channels on \emph{all} of the isotopes of W are positive and in the 5-10~MeV range. The extra energy produced by these exothermic reactions is available to the \(\alpha\) particle when it is emitted. This is combined with the energy it already gains from the momentum of the absorbed neutron, which for a 14~MeV neutron on \(^{184}\)W, for example, could be as much as 13.7~MeV by a simple mass distribution argument (with the \(^{181}\)Hf residual taking the remainder). With a Q-value of 7.3~MeV for the \(^{184}\)W\((n,\alpha)\)\(^{181}\)Hf in the TENDL-2014 library, this then explains how the \(\alpha\) particle can have emission energies of 20~MeV or more, especially given the fact the maximum Q-value for \((n,\alpha)\) on W is 9.1~MeV on \(^{183}\)W.

This important observation about the energy of light particle emissions, which is not well known or appreciated in the literature, is also observed to a lesser extent in Fe. While the Q-value for the \((n,\alpha)\) on \(^{56}\)Fe is only 325.9 keV, for the the heavier \(^{57}\)Fe it is 2.4~MeV, and for the lighter \(^{54}\)Fe 0.8 MeV, explaining why there is a small chance of producing \(\alpha\) particles with energies around 14~MeV.

Even though the probability of production for these high-energy \(\alpha\) particles, and to a lesser extent protons, is low compared to the dominant scattering channels (particularly in W), the extreme nature of the energies they are produced at suggests that they cannot be ignored when considering the accumulation of radiation damage.

\begin{figure}
\includegraphics[width=1.0\textwidth]
{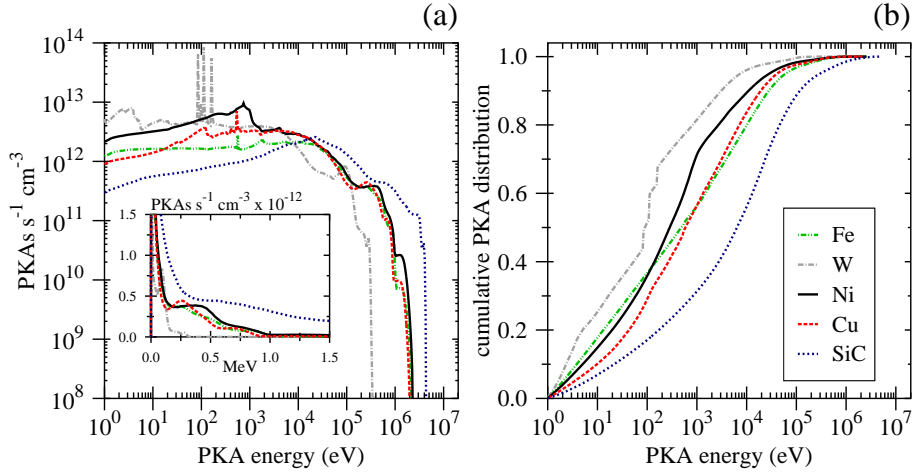}
\caption{\label{elemental_comp} (a) Total PKA spectra and (b) cumulative PKA distributions for different elements under the DEMO hcpb FW conditions. Note that in (b), and in other cumulative curves presented later, the distributions begin from 1~eV. The conversion to PKAs~s\(^{-1}\)~cm\(^{-3}\) was performed using the densities and masses given in Table~\ref{demohcpbfw}, where the total PKAs and above 10~eV average PKA energies for each distribution are also presented.}
\end{figure}

Total PKA spectra for several materials under the same DEMO hcpb FW neutron spectrum are shown in Fig.~\ref{elemental_comp}a, illustrating one of the benefits of such a simplification -- the ability to compare the distribution of initial damage events between different materials (or indeed different irradiation scenarios -- see later). Note that for the Carbon in SiC, the nuclear data file used in NJOY is that of natural C rather than a combination of \(^{12}\)C and \(^{13}\)C. In TENDL-2014~\cite{tendl2014} this is the only data for C and comes directly from JENDL-4.0~\cite{jendl4.0}, which is not as complete -- with missing or erroneously merged reaction channels and energy-angular distributions -- compared to the data for other nuclides.

For the total spectra in Fig.~\ref{elemental_comp} the contribution from PKAs of light particles have been omitted (via an option within the code). The damage produced by these high-energy, low mass particles is completely different to similarly high-energy heavy particles and so it would be misleading to include them in total spectra for the materials considered in the figure (even for SiC). Simulations using the binary collision approximation (BCA) performed with the SRIM (Stopping and Range of Ions in Matter) code in full cascade mode (see~\cite{srim} for more details of SRIM) for 14~MeV \(^4\)He ions impinging on Fe and W reveal that the average energy of the above 10~eV (primary) recoils generated by the helium ions is only 76 and 118~eV, respectively. Thus the distribution of recoils from \(\alpha\) particles, which are essentially the secondary knock-on atoms (SKAs) relative to the original incident neutron, would produce primarily single point defects or very small clusters, with very occasional extended defect regions. In comparison, the above 10~eV average energies for recoils generated by equivalent 14~MeV self-ions is 960~eV and 8~keV (again calculated in SRIM), respectively, in Fe and W, which would produce very much greater defect sizes and distributions. Thus it is inappropriate to have heavy ion recoils at the energies associated with the light particle recoil energies -- if the damage produced by such light particles is considered significant then the best option would be to evaluate the SKA spectrum, for example with SRIM, and then add this to the total PKA spectrum from heavy particles.

Another useful representation for modelling is the cumulative distribution of PKAs, such as those shown in Fig.~\ref{elemental_comp}b, for the same total PKA spectra in Fig.~\ref{elemental_comp}a. Here PKA events with energies less than 1~eV are omitted, since these do not contribute to damage production. These cumulative distributions could be used to statistically sample PKA events, for example in a Monte-Carlo calculation, or in a stochastic cluster dynamics simulation (see Section~\ref{scd_section}). The sampling rate, i.e. the PKAs~s\(^{-1}\), per unit volume is obtained by integrating (above 1~eV here) the equivalent total PKA curve from Fig.~\ref{elemental_comp}a, and are given for these curves in Table~\ref{demohcpbfw}. As would be predicted, the PKAs in a heavy material like W are, on average, lower in energy than in lighter materials such as Fe. Indeed, the average PKA energy can be evaluated directly, and the results for these total PKA distributions are also shown in Table~\ref{demohcpbfw}. Since the characteristic threshold displacement energy, below which a PKA will not escape from its lattice site, is of the order of \(10^1\)~eV, there is no point including PKA energies below this energy in the averaging. Thus all of the average PKA energies shown in Table~\ref{demohcpbfw} and elsewhere are for PKAs strictly above 10~eV in energy. From the table we see that W has a very low average of only 3.3~keV, with the average in Fe significantly higher at 19.0~keV.

\begin{table}
\caption{\label{demohcpbfw} Total PKAs (above 1~eV) and average PKA energy (above 10~eV) for various materials caused by the FW conditions in the hcpb DEMO reactor (Fig.~\ref{spectra}). The calculated molar masses (using the 2012 atomic mass evaluations~\cite{ame2012} with the isotopic abundances for each material) and assumed material densities, used to convert to PKAs per unit volume, are also given.}
%\begin{tabular}{cr@{.}lr@{.}lccc}
\begin{tabular}{ccp{0.7cm}r@{.}lD..{5.3}D..{5.3}}\hline\hline\vspace{-0.25cm}\\
Material& Total PKAs above 1~eV & \multicolumn{3}{c}{Average PKA energy}
& \multicolumn{1}{c}{Molar Mass}& \multicolumn{1}{c}{Density}\\
&(PKAs~s\(^{-1}\)~cm\(^{-3}\))& \multicolumn{3}{c}{above 10~eV (keV)}
&\multicolumn{1}{c}{(amu)}& \multicolumn{1}{c}{(g~cm\(^{-3}\))} \\\hline
Fe & 4.33E+14 && 18 & 8 & 55.85 & 7.9 \\
W & 1.11E+15 && 3 & 2 & 183.84 & 19.3 \\
Ni & 9.48E+14 && 10 & 4 & 58.69 & 8.9 \\
Cu & 5.77E+14 && 13 & 2 & 63.55 & 9.0 \\
SiC & 3.16E+14 && 76 & 1 & 20.05 & 3.2 \\
\hline
\end{tabular}
\end{table}

Interestingly, these values of average PKAs are significantly lower than typical values presented in the literature, where, for example, the average PKA energy was thought to be about 150~keV for W under fusion neutron irradiation~\cite{sand2013}. This large discrepancy comes from the assumptions made in other calculations. The present results have been calculated for a full FW fusion neutron spectrum (Fig.~\ref{spectra}), with its significant proportion of moderated, lower energy neutrons in addition to the main 14~MeV peak and, furthermore and more importantly, the complete nuclear reaction cross-sections, including full angular dependencies, have been used to produce the recoil cross-section matrices with NJOY. The literature values, on the other hand, are obtained via an approximation of the neutron fields to a single delta-function at 14~MeV, and only the scattering channel with an isotropic angular distribution is considered. Fig.~\ref{Fe56_matrix_2d} demonstrates that this last approximation is not sufficient as there is significant deviation from isotropy (an isotropic response would have produced constant flat distributions for the distributions shown there). This observation has important implications for radiation damage modelling and will be discussed further in a future publication.

\section{Variations in time and space}

The PKA spectra for the complete set of naturally occurring elements (excluding actinides) under DEMO FW conditions have recently been calculated for inclusion in a comprehensive nuclear physics handbooks~\cite{handbook2015}, which also describes the neutron-induced transmutation and activation of material. Elemental PKA distributions, in particular, are presented in great detail, with an additional supplement~\cite{handbooksupplement2015} giving the distributions, such as those in Fig.~\ref{elemental}, in tabular form. However, with the methodology developed here, it is a simple matter to consider the variation in PKA spectra due to different irradiation scenarios, perhaps from different positions within a nuclear reactor model.

\begin{figure}
\includegraphics[height=0.5\textwidth]
{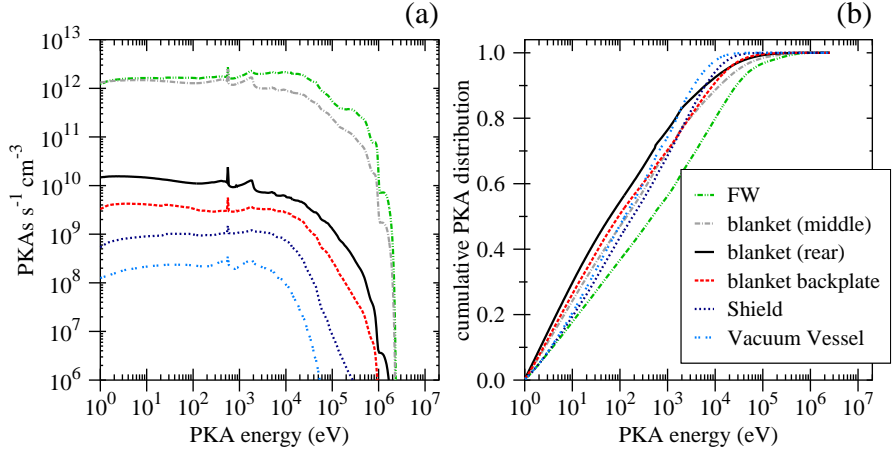}
\caption{\label{space_comp} (a) Total PKA spectra and (b) cumulative PKA distributions for pure Fe as a function of position (depth) within the outboard equatorial wall of the hcpb DEMO concept. The conversion to PKAs per unit volume was performed using the density and molar mass of pure Fe given in Table~\ref{demohcpbfw}. The total PKAs and above 10~eV average PKA energies for each distribution are given in Table~\ref{time+space}. }
\end{figure}

Fig.~\ref{space_comp} shows the total PKA spectra (a) and cumulative PKA distributions (b) for pure Fe as a function of position into the equatorial wall of the hcpb DEMO reactor vessel, while the total PKAs above 1~eV and PKA average energies (above 10~eV) are given in Table~\ref{time+space}. In order of depth the different regions considered are: the FW itself (0-2.3~cm depth from plasma face), as shown in Fig.~\ref{spectra}; the middle and rear of the tritium breeding blanket (at around 14 and 77~cm depth respectively); the blanket backplate (\(\sim\)80-110~cm); the vacuum vessel (VV) shield (115-160~cm); and the VV itself (around 2~m depth). In agreement with the decrease in total neutron flux, as a function of depth the total number of PKAs in pure Fe decreases dramatically, with the above 1~eV PKAs falling by 4 orders of magnitude between the FW and VV, from \(4.3\times10^{14}\)~PKAs~s\(^{-1}\)~cm\(^{-3}\) to only \(4.3\times10^{10}\)~PKAs~s\(^{-1}\)~cm\(^{-3}\). However, from Fig.~\ref{space_comp}a it is also evident that the proportion of PKAs at higher energies is decreasing faster with depth than the overall PKA levels, which is confirmed by the drop in the above 10~eV average PKA energy (Table~\ref{time+space}). On the other hand, the cumulative PKA distributions shown in Fig.~\ref{space_comp}b demonstrate that the picture is slightly more complex than this. While it is the case that through the FW and blanket the PKA distributions shift to lower and lower energies, beyond this there is not such a clear trend. Indeed, the blanket backplate has a slightly higher proportion of higher-energy PKAs than the rear of the blanket itself. There is a subtle interplay between moderation and overall neutron absorption that produces this changing picture, with initially (near the plasma) high moderation levels, but relatively little absorption, before, once a significant proportion of neutron have been moderated, increased absorption of these moderated neutrons in particular.

\begin{figure}
\includegraphics[height=0.5\textwidth]
{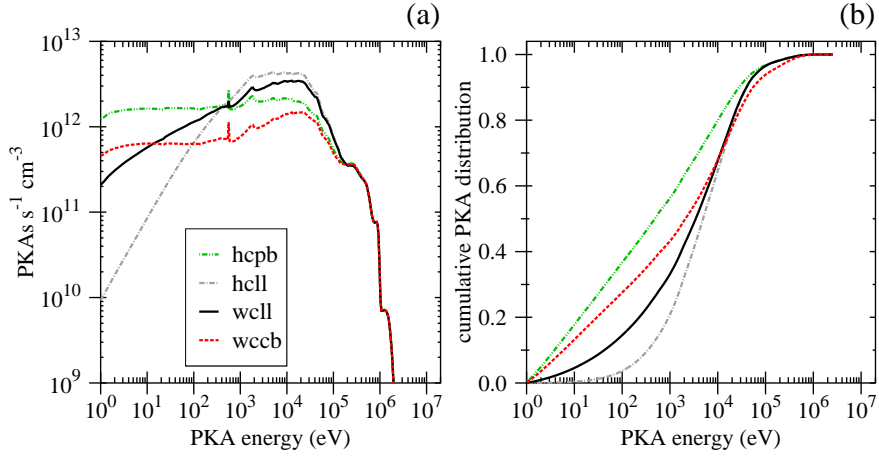}
\caption{\label{reactor_comp} (a) Total PKA spectra and (b) cumulative PKA distributions for pure Fe in the equatorial FW of a DEMO fusion reactor as a function of the cooling and tritium breeding concept. See the main text for details of the designations of the different concepts. The total PKAs and above 10~eV average PKA energies for each distribution are given in Table~\ref{time+space}.}
\end{figure}

We can also compare the PKA spectra between different reactors, as shown in Fig.~\ref{reactor_comp} for the equatorial FW of the four different DEMO concepts discussed in~\cite{gilbertetalfst2014}: the helium-cooled reactor with a tritium-breeding blanket of Li+Be pebbles already introduced (hcpb); a helium-cooled reactor with a self-cooling liquid LiPb blanket (hcll), a water-cooled reactor with the liquid blanket (wcll), and a water-cooled Li+Be ceramic breeder concept (wccb). The neutron irradiation field for FW position within each concept is shown in Fig.~\ref{spectra}. The results here indicate that although for the wccb concept the total PKA rate is the lowest (see Table~\ref{time+space}), the average PKA energy above 10~eV is actually significantly higher at 32.0~keV compared to the other three concepts that all have averages around 19-21~keV.

Finally, the present methodology can be combined with inventory calculations, to investigate how the PKA spectra for a material might change with time due to transmutation-induced changes in chemical composition. Using the inventory code FISPACT-II~\cite{subletetal2012}, together with the modern TENDL~\cite{tendl2014} nuclear data libraries we can compute the variation in composition in pure W under the fission spectrum evaluated for the High Flux Isotopes Reactor (HFIR) at Oakridge National Laboratory. The neutron spectrum, reproduced from~\cite{greenwood1994}, is shown in Fig.~\ref{spectra}. This spectrum, as calculated and presented in the literature, has a very high proportion of low energy neutrons, leading to very high transmutation rates (particularly in W), which is useful here to illustrate the concept. Indeed after only 5 years of exposure to the HFIR neutron field, the initially pure W material contains more than 50 atomic \% of transmutation products different from W, including Os, Pt, Re, and Ir.

\begin{figure}
\includegraphics[height=0.5\textwidth]
{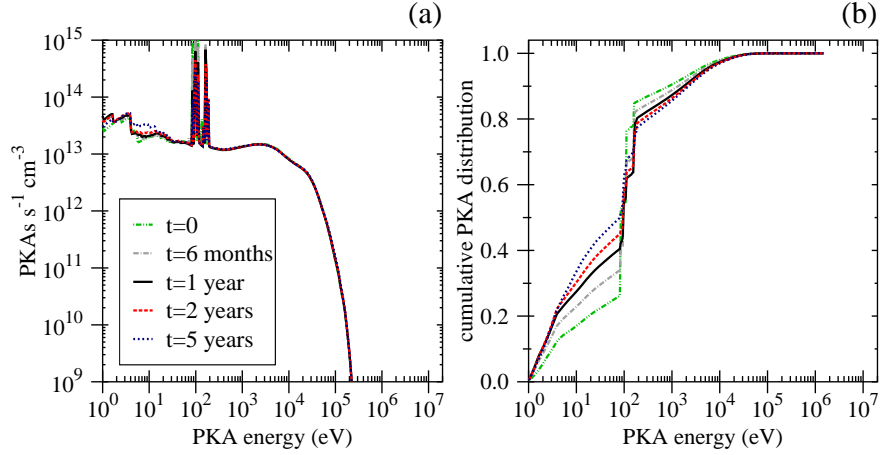}
\caption{\label{time_comp} (a) Total PKA spectra and (b) cumulative PKA distributions for initially pure W as a function of total exposure time to a typical neutron spectrum for the HFIR test reactor. The total PKAs and above 10~eV average PKA energies for each distribution are given in Table~\ref{time+space}.}
\end{figure}

Fig.~\ref{time_comp} shows how the total PKA spectrum and cumulative distribution varies in the material as a function of time. These curves have been obtained from the \texttt{SPECTRA-PKA} code with recoil cross-section matrices calculated by NJOY using the same TENDL libraries as used in the inventory calculations. For computational reasons, and also because low concentration nuclides do not contribute significantly, only nuclides making up more than 0.1 atomic \% of the composition at each time were included in the PKA calculations. The total PKA rates and energy averages at each time are also given in Table~\ref{time+space}.

At first glance, the total PKA spectra in Fig.~\ref{time_comp}a do not look all that different. However, Table~\ref{time+space} shows that the total PKAs (above 1~eV) is varying quite significantly, dropping by almost 40\% between \(t=0\) and \(t=5\)~years, while, at the same time, the average PKA energy increases by roughly the same fraction. Note that the material density and molar mass assumed for all compositions, and used to calculate the  PKAs~s\(^{-1}\)~cm\(^{-3}\), was the same as that in pure W (see Table~\ref{demohcpbfw}). In reality, both the density and molar mass change slightly with time, but the difference is not significant enough to produce a visible change in the PKA distributions.

\begin{table}
\caption{\label{time+space}Total PKAs (above 1~eV) and average PKA energy (above 10~eV) for Fe in different fusion DEMO irradiation fields, and for W as a function of time in the HFIR experimental reactor.}
%\begin{tabular}{cr@{.}lr@{.}lccc}
\begin{tabular}{clcp{0.7cm}r@{.}l}\hline\hline\vspace{-0.25cm}\\

Element&Conditions& Total PKAs above 1~eV & \multicolumn{3}{c}{Average PKA energy}
\\
&&(PKAs~s\(^{-1}\)~cm\(^{-3}\))& \multicolumn{3}{c}{above 10~eV (keV)}
\\\hline
Fe & FW hcpb & 4.33E+14 && 18 & 8 \\
Fe & blanket (middle) hcpb & 3.00E+14 && 10 & 3 \\
Fe & blanket (rear) hcpb & 2.58E+12 && 6 & 0 \\
Fe & blanket backplate hcpb & 7.83E+11 && 5 & 5 \\
Fe & Shield hcpb & 2.03E+11 && 2 & 8 \\
Fe & Vacuum Vessel hcpb & 4.31E+10 && 1 & 6 \\

Fe & FW hcll& 4.48E+14 && 20 & 5 \\
Fe & FW wcll& 4.21E+14 && 20 & 8 \\
Fe & FW wccb& 2.21E+14 && 31 & 5 \\

W & HFIR t=0& 8.82E+15 && 1 & 0 \\
W & HFIR t=6 months& 7.30E+15 && 1 & 2 \\
W & HFIR t=1 year& 6.43E+15 && 1 & 5 \\
W & HFIR t=2 years& 6.00E+15 && 1 & 6 \\
W & HFIR t=5 years& 5.68E+15 && 1 & 8 \\

\hline
\end{tabular}
\end{table}

\section{Application to radiation damage}
Ideally, PKA distributions such as those presented here would be used in combination with detailed atomistic simulations of displacement cascades to predict the production of damage, perhaps followed by Monte-Carlo approaches to trace the long-term evolution of large distributions of defects. Describing the defect formation in displacement cascades and subsequent evolution is currently receiving significant attention, with several groups working on using a combination of experimental observations and atomistic simulations, particularly for W~\cite{sand2013,yietal2015}, to define defect clustering and size-scaling laws. However, while those works are still maturing, here we consider simpler approximations that can make use of the PKA spectra presented in the present work.

Firstly, the Frenkel pair (FP) production rate can be approximated by combining elemental PKA spectra, such as those in Fig.~\ref{elemental} for the DEMO FW, with distributions as a function of ion energy (\(E_{\rm ion}\)) of the FPs-per-ion (\(N_{\rm FPs}\)) calculated using the a suitable binary collision approximation (BCA) code -- and for the present the well-known Monte-Carlo based SRIM\cite{srim} program has been used. For each PKA species predicted for neutron irradiated Fe and W (i.e. Fe, Cr, Mn, He, and H into Fe, and W, Ta, Hf, He, and H into W - see Fig.~\ref{elemental}) a range of \(E_{\rm ion}\) values were simulated, covering the range of predicted PKA energies. For each \(E_{\rm ion}\), 1000 ions were simulated in SRIM's full cascade mode to produce a statistically reliable number of vacancies per ion, which is equivalent, in this approximation, to \(N_{\rm FPs}\). For the threshold displacement energy \(E_d\), which controls the minimum energy required to produce a stable FP, standard literature values (see, for example, table II in~\cite{macfarlanekahler2010}) of 40~eV and 90~eV were used for Fe and W, respectively. The resulting \(N_{\rm FPs}\) versus \(E_{\rm ion}\) curves were then fitted to an appropriate functional form:
\begin{equation}\label{srimfunction}
N_{\rm FPs}(E_{\rm ion})=A\ln(BE_{\rm ion}+C)+DE_{\rm ion}+F,
\end{equation}
where \(A, B, C, D\), and \(F\) are the parameters to be fitted. The fitting itself was performed using the in-built capabilities of the GNUPLOT~\cite{gnuplot5} data plotting program. Table~\ref{srimfits} gives the values of these fitted parameters for the ten ion implantation types necessary for the simulation of PKAs in Fe and W.

\begin{table}
\caption{\label{srimfits} The fitted parameters of Eq. \eqref{srimfunction} for different ion species implanted into either Fe or W using SRIM\cite{srim}.}
%\begin{tabular}{cr@{.}lr@{.}lccc}
\begin{tabular}{c|ccccc}\hline\hline\vspace{-0.25cm}\\
Ion implantation event & \(A\) & \(B\) & \(C\) & \(D\) & \(F\) \\
\hline
 Fe into Fe &3.21E+03  &2.70E-04  &9.98E-01  &5.07E-01  &7.99E+00 \\
 Cr into Fe &1.01E+04  &1.46E-04  &1.01E+00  &-1.84E-01  &-5.40E+01 \\
 Mn into Fe &3.83E+03  &2.53E-04  &9.98E-01  &4.11E-01  &7.36E+00 \\
 He into Fe &2.15E+01  &2.30E+01  &-5.08E-01  &3.68E-01  &2.04E+01 \\
 H into Fe &2.60E+00  &4.97E+00  &2.20E-01  &4.42E-01  &3.83E+00 \\
 \hline
 W into W &4.29E+02  &4.33E-04  &1.00E+00  &4.34E-01  &3.02E-01 \\
 Hf into W &7.75E+02  &2.82E-04  &1.05E+00  &3.98E-01  &-3.28E+01 \\
 Ta into W &6.50E+02  &2.91E-04  &1.00E+00  &4.36E-01  &9.96E-01 \\
 He into W &1.01E+01  &7.43E+00  &5.05E-01  &2.57E-01  &6.32E+00 \\
 H into W &3.31E+00  &4.69E-01  &8.44E-01  &1.50E-01  &5.76E-01 \\ 
\hline\hline
\end{tabular}
\end{table}

\begin{figure}
\includegraphics[width=0.5\textwidth]
{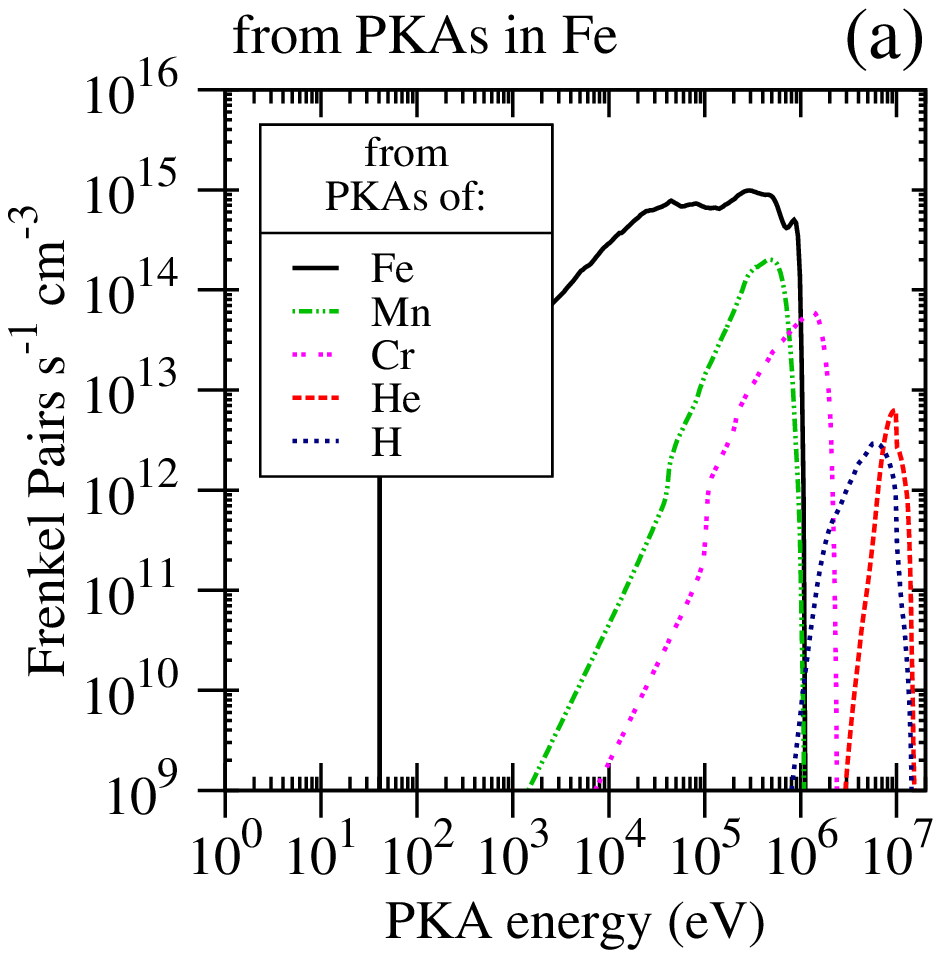}
\includegraphics[width=0.5\textwidth]
{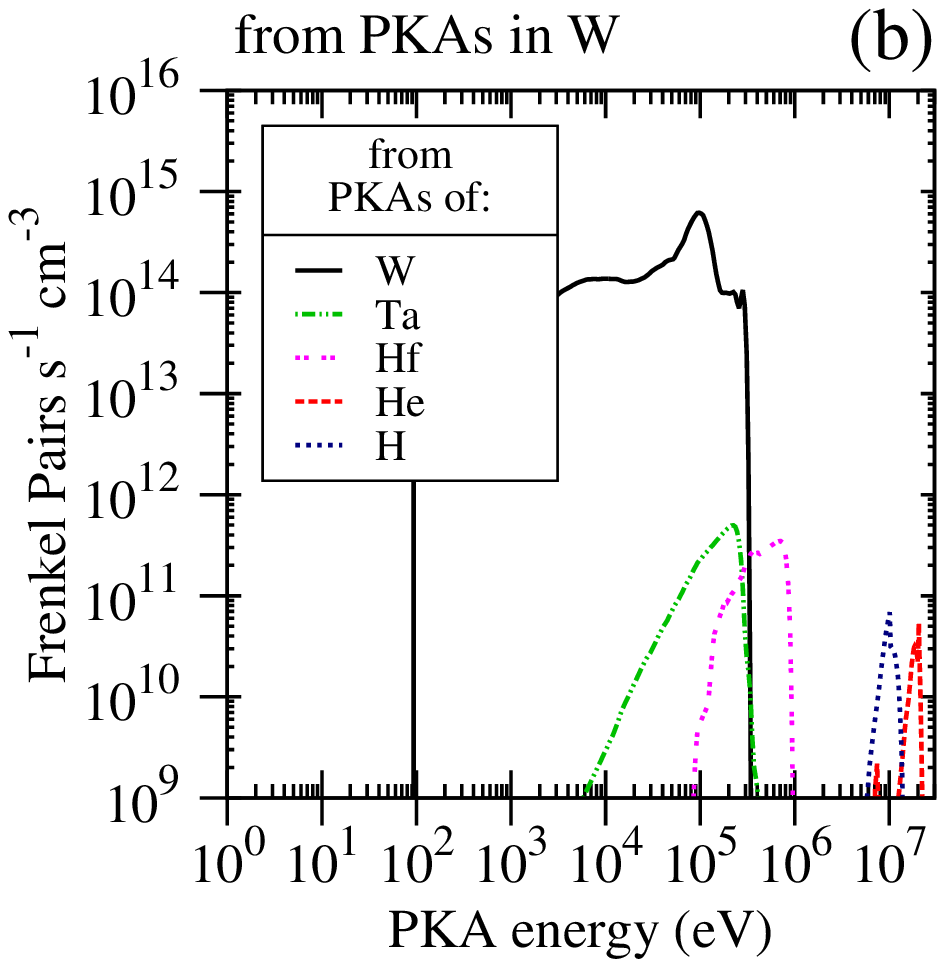}
\caption{\label{SRIM_FPs} The Frenkel pair production rates for (a) Fe and (b) W under the DEMO hcpb FW neutron irradiation scenario predicted using the BCA code SRIM~\cite{srim}. The plot shows the curves for each of the different PKA distributions shown in Fig.~\ref{elemental}.}
\end{figure}

Fig.~\ref{SRIM_FPs} shows the FPs~s\(^{-1}\)~cm\(^{-3}\) distributions obtained by combining the fitted FP curves with each of the elemental PKA curves from Fig.~\ref{elemental}. Note that for both Fe (Fig.~\ref{elemental}a) and W (Fig.~\ref{elemental}b) only the FP distributions produced by PKAs of the original host element have a lower limit at the \(E_d\) values used in the SRIM calculations. While these values of \(E_d\), chosen somewhat arbitrarily from the literature, influence the results for all PKA species as far as absolute FP numbers are concerned, it is only the self-PKA distributions that have energies in the 10s of eV range or below (due to neutron scattering) and are thus truncated by \(E_d\).
Table~\ref{demohcpbfwFPs} gives the total FP production rate from each PKA species, together with overall total FP production rate for the irradiated Fe and W. The table and figure demonstrate the expected dominance of the damage produced by the host element (Fe or W), particularly in the case of W, where the FPs produced by the W PKAs represent more than 99\% of the \(6.54\times10^{19}\)~FPs~s\(^{-1}\)~cm\(^{-3}\).

\begin{table}
\caption{\label{demohcpbfwFPs} Frenkel pair production rates in Fe and W from the different PKA species generated under the neutron field predicted in the FW of a conceptual design of a hcpb DEMO fusion reactor.}
%\begin{tabular}{cr@{.}lr@{.}lccc}
\begin{tabular}{c|ccccc|c}\hline\hline\vspace{-0.25cm}\\
Element&\multicolumn{6}{c}{Frenkel Pair production rates (FPs~s\(^{-1}\)~cm\(^{-3}\))}
\\\hline&\multicolumn{5}{c}{Frenkel Pairs from:}\vline&\multirow{2}{*}{Total}\\
& Fe & Mn & Cr & He & H\\
Fe & 6.48E+20 & 8.66E+19 & 6.55E+19 & 2.20E+19 & 1.56E+19 &  8.38E+20
\\\hline&\multicolumn{5}{c}{Frenkel Pairs from:}\vline&\multirow{2}{*}{Total}\\
& W & Ta & Hf & He & H\\
W& 6.25E+19 & 7.83E16& 1.82E17 &1.75E17 & 1.69E17 &6.31E19 \\

\hline\hline
\end{tabular}
\end{table}

Confirming the earlier assertion (see Section~\ref{realmaterials}) the figure shows that the damage, in this case measured as theoretical FP production, caused by the high energy light protons and \(\alpha\) particles is indeed low compared to the heavier PKA species, even accounting for the reduced probability of production. For example, in Fig.~\ref{elemental}a, the peak PKA rate from Mn is similar to that of the protons produced from the same reaction channels on Fe. The protons, however, are mainly generated in the 5-10~MeV range compared to the few 100s of keV range of the Mn, but nonetheless produce a peak FP rate that is at least two orders of magnitude lower than that from Mn. This observation justifies the omission of the light particle PKAs from the total PKA curves discussed earlier, and that the \(\alpha\)/proton distributions should be used to generate appropriate SKA distributions that can then be included in the total PKA distributions.

Fig.~\ref{SRIM_FPs}, for Fe and W, also illustrates the relative importance of different PKA energies to `damage' production. Earlier, it was observed that the average PKA energy for Fe and W under the DEMO FW conditions was less than 20~keV (see Table~\ref{demohcpbfw}). However, after multiplying by a damage measure, in this case the FP production predicted by SRIM, it is clear that the important range of PKAs is much higher than this. In Fig.~\ref{SRIM_FPs}a, the dominant FP~s\(^{-1}\)~cm\(^{-3}\) curve from Fe PKAs has a maximum of almost \(1\times10^{15}\)~FP~s\(^{-1}\)~cm\(^{-3}\) at around 300~keV, with a broad range of PKA energies -- from 10 to 900~keV -- giving FP production rates above \(1\times10^{14}\)~FP~s\(^{-1}\)~cm\(^{-3}\). Thus, when performing modelling and simulation of damage creation and evolution, a large range of PKA energies must be considered.

On the other hand, in W, Fig.~\ref{SRIM_FPs}b, indicates that there is instead a very pronounced peak in damage due to PKAs at approximately 100~keV in energy, which is very different from the 3.2~keV average obtained from the raw PKA curves.

\subsection{Stochastic cluster dynamics}\label{scd_section}

The BCA calculations discussed above are a very rough approximation of damage produced under irradiation, and in particular do not take into account the evolution of defect populations, via migration, recombination and clustering. An improved approach that attempts to include some aspects of this behaviour was investigated by Marian \etal~\cite{marianbulatov2011,marianhoang2012}, who have employed the stochastic cluster dynamics method to evaluate irradiation damage accumulation on pure Fe~\cite{marianbulatov2011} and W~\cite{marianhoang2012} including the effect of He and H.

Here we use SCD to explore defect clustering during the initial stages of irradiation exposure of Fe and W in the same DEMO FW spectrum considered above. The SCD computational methodology for solving the stochastic variants of the mean-field rate theory ODEs described in~\cite{marianbulatov2011} has been modified to use directly the PKA source terms calculated in the present work, rather than approximating the PKA production from the neutron spectrum. For this purpose, the SCD code samples the cumulative PKA distributions of heavy recoils for Fe and W shown in Fig.~\ref{elemental_comp}b (\ie without considering the implications of different recoil species) at the total PKA rates given in Table~\ref{demohcpbfw}. Once a PKA energy \(E_{\rm PKA}\) is chosen, the number of FPs created, \(N\), is governed by a non-linear damage production law that reflects the deviation from the NRT formalism as given by molecular dynamics (MD) analysis. This law is a simple power law for W:
\begin{equation}
N=aE_{\rm PKA}^b,
\end{equation}
or an exponential correction on the NRT linear relationship for Fe:
\begin{equation}
\begin{split}
N=&\eta\frac{kE_{\rm PKA}}{2E_{th}},\\
\eta=&\exp\left(-cE_{\rm PKA}\right)
\end{split}
\end{equation}
Here \(a\), \(b\), and \(c\) are adjustable parameters defined via statistics extracted from MD simulations of displacement cascades. For Fe, $c=3.57$ as given by Malerba \etal~\cite{malerba2006}, and for W, $a=1.49$ and $b=0.82$ \cite{troevetal2011} ($E_{\rm PKA}$ in keV). Clusters of various sizes are then inserted according to a binomial distribution that obeys the global defect cluster fractions until \(N\) is exhausted. In W, these fractions are 0.5 and 0.2 for self-interstitial atoms (SIAs) and vacancies, respectively~(Fikar and Sch\"aublin~\cite{fikarschaublin2009}), while in Fe they are 0.55 and 0.25 \cite{malerba2006}. Note also that there are hard limits of 0.62 (W) and 0.33 keV (Fe) in each case \cite{troevetal2011,malerba2006} below which a PKA does not produce any damage.

Alongside this introduction of clusters directly in cascades, the populations evolve via diffusion, which varies as a function of cluster size and nature, via binary reactions, by recombination, and by annihilation at sinks. In the latter case, dislocations and grain boundaries are considered inexhaustible defect sinks~\cite{marianhoang2012}. The sink strengths in each case depend on the grain size and dislocation densities considered. For W, these were 100 $\mu$m and $10^{14}$ m$^{-2}$, while for Fe they were 50 $\mu$m and $1.5\times10^{15}$ m$^{-2}$ (see \cite{marianbulatov2011} and \cite{marianhoang2012} for details).

To promote vacancy clustering and SIA survival in this formulation we follow~\cite{marianhoang2012}, and insert He into the system at rates calculated by the inventory simulator FISPACT-II~\cite{subletetal2012} for each material under the same neutron irradiation conditions. For the DEMO FW scenario considered here the calculated rates are \(3.6\times10^{-6}\) and \(6.7\times10^{-6}\)~appm~He~s\(^{-1}\) for Fe and W, respectively, which are the average values during 1 full power year (fpy) of exposure. Here appm is atomic parts per million. At this point, hydrogen is not considered due to a lack of a full data set, although the SCD formulation can treat another species trivially by construction.

\begin{figure}[htpb]
%{\bf Latest -- with He}\\
\includegraphics[height=0.5\textwidth]
{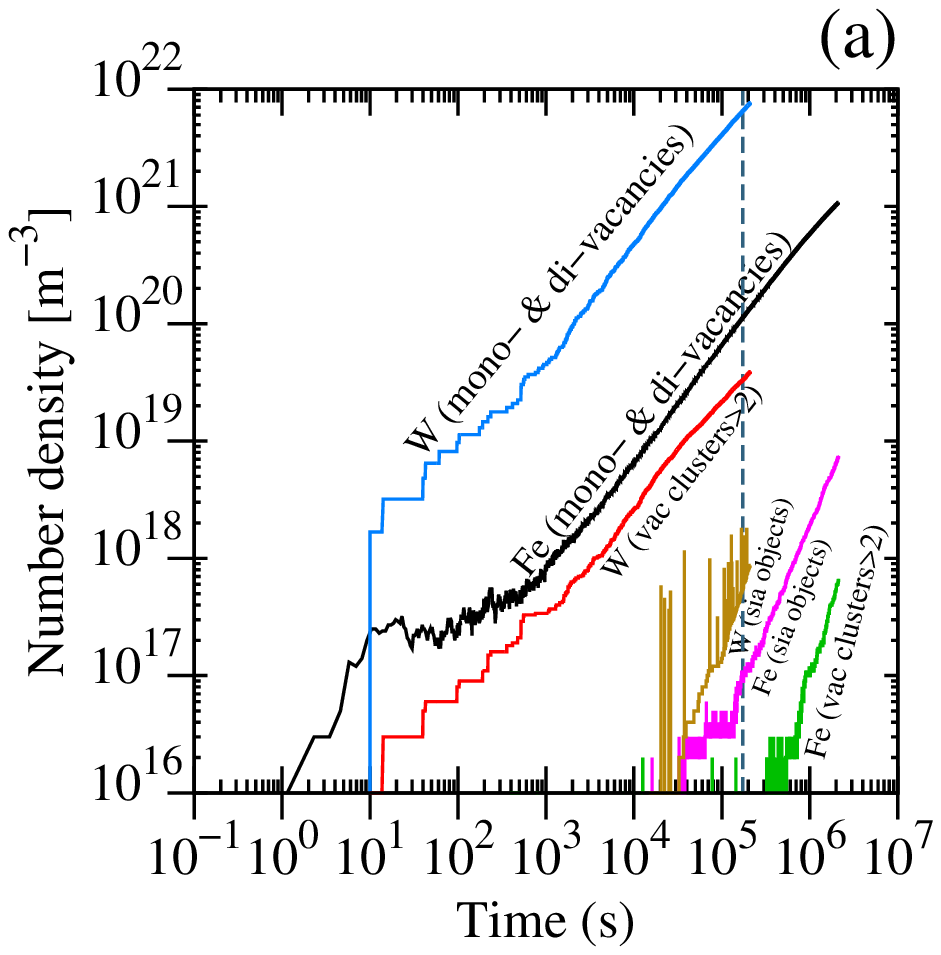}
\includegraphics[height=0.5\textwidth]
{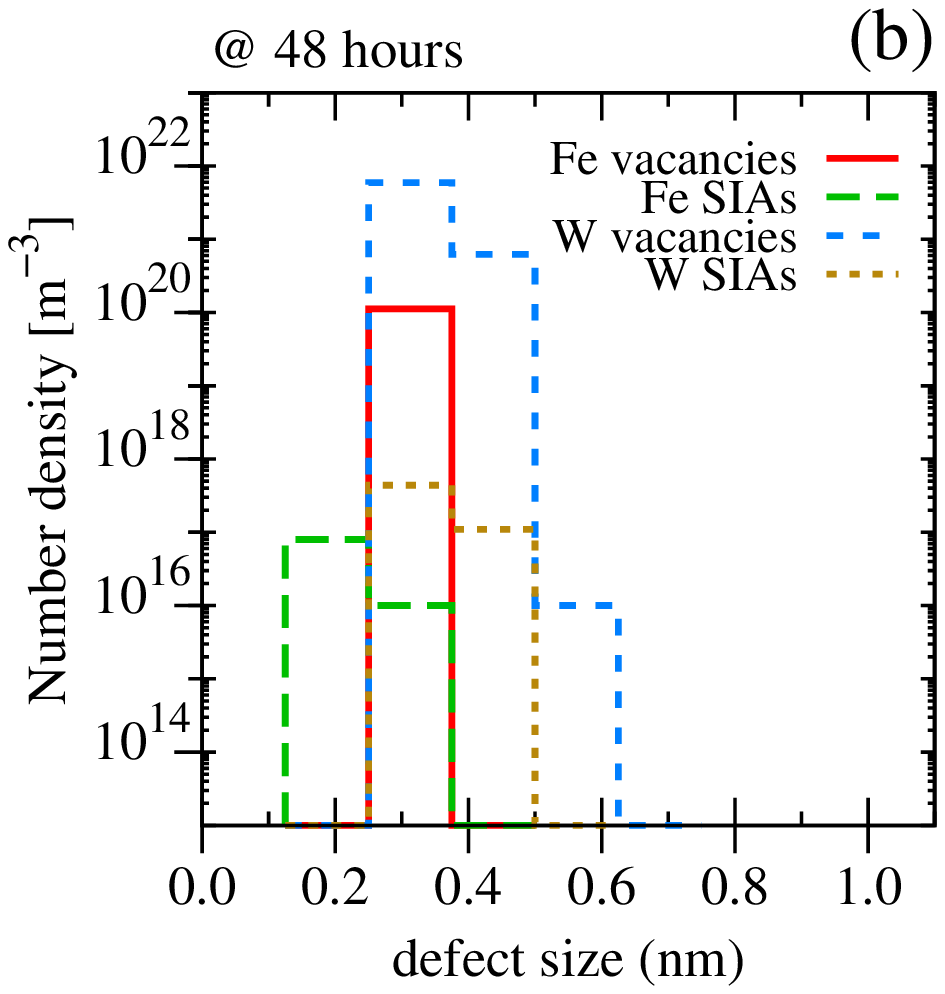}
\caption{\label{SCD_fig} SCD simulation results. (a) the evolution in concentration of small point-like vacancy defects (mono- and di-vacancies), larger vacancy clusters, and SIA clusters of any size as a function simulation time in Fe and W. (b) The size distribution of vacancy and SIA clusters after simulating for 48 hours of real time. The dashed vertical line in (a) corresponds to this 48-hours of simulation. In (b) it is assumed that vacancy clusters adopt a spherical shape and that SIAs are arranged as circular loops~\cite{marianhoang2012}, with the size referring to the diameter of the corresponding spherical or circular shape.}
\end{figure}

Fig.~\ref{SCD_fig}a shows the defect accumulation results in Fe and W during simulation at a temperature of 300K, which was chosen to be in agreement with the effective temperature employed when calculating the nuclear recoil cross-sections with NJOY and the neutron-energy spectrum for the DEMO FW with MCNP -- see Section~\ref{methodology}. In the figure, the vacancy populations, are split into those that are essentially immobile (more than 2 vacancies per cluster) and the mobile point vacancy defects, comprising the sum of mono- and di-vacancies. Note that the simulation volume assumed in these simulations was \(10^{-10}\)~cm\(^3\) and hence the minimum concentration -- when there is one defect in a given population -- corresponds to the \(10^{16}\)~m\(^{-3}\) minimum in Fig.~\ref{SCD_fig}a (this limit has been relaxed in the Fig.~\ref{SCD_fig}b bar graph to give a proper sense of the concentrations at different sizes).

In contrast to the BCA results, where Fe was observed to have the highest FP production rate, the SCD simulations suggest that W will experience the greater defect production rate, with, for example, at least an order of magnitude more mobile vacancy defects after the same simulation time.

The initial annihilation and recombination rate in Fe is much higher than in W, partially due to the differential mobility of defects, which -- even at this modest temperature -- can be of two orders of magnitude.  Furthermore, the combined effect of the smaller grain size and higher dislocation density in Fe also increases defect annihilation. The overall result is a delay in the formation and accumulation of defects in Fe. This is further illustrated by comparing the size distribution of defects. For example, after 48 hours (Fig.~\ref{SCD_fig}b) both vacancy and SIA defects in Fe have a significantly smaller size distribution than the equivalent populations after the same simulation time in W. Note that for these size distributions we have assumed that vacancies adopt a spherical shape and that SIAs are arranged as circular loops~\cite{marianhoang2012}, with the size referring to the diameter of the corresponding spherical or circular shape.

Even though the defect evolution modelled here is not the complete picture, it nonetheless demonstrates the importance of including evolution when attempting to quantify and compare radiation damage in materials. Neglecting this evolution in Fe and W, as was the case in the previous section when using a damage function from BCA calculations, suggested an entirely opposite comparison between these two important fusion materials.

%\begin{figure}[htpb]
%{\bf ORIGINAL -- no He}\\
%\includegraphics[height=0.5\textwidth]
%{figure12a_first}
%\includegraphics[height=0.5\textwidth]
%{figure12b_first}
%\caption{\label{SCD_fig_first} SCD simulation results. (a) the evolution in concentration of small point-like vacancy defects %(mono- and di-vacancies) and larger vacancy clusters as a function simulation time in Fe and W. (b) The size distribution of %vacancy clusters (of all sizes) after simulating for 3 hours of real time. In (b) the vacancies are assumed to adopt a %spherical shape and the size refers to the diameter of the spherical voids.}
%\end{figure}

\section{Summary}

A detailed understanding of the initial knock-on events generated under neutron irradiation is a vital component in the investigation of radiation damage accumulation. The methodology described in this paper for predicting the spectra of primary knock-on atoms (PKAs) produced when a material is subjected to a field of neutrons should make their calculation a straightforward and routine part of materials research for nuclear applications.

Using group-wise incident-to-recoil energy cross-section matrices processed from the latest, modern and complete nuclear data libraries by the NJOY-12~\cite{njoy2012} system, a newly written code \texttt{SPECTRA-PKA} collapses this data with a neutron irradiation field of interest, and produces the PKA spectrum for every possible nuclear reaction channel on a target nuclide or sequence of targets. In particular, the unique ability to merge and sum data for several target nuclides enables the PKA distributions in complex (real) materials subjected to non-trivial neutron fields to be evaluated.

\texttt{SPECTRA-PKA} performs post-processing of the raw PKA spectra separated as a function of nuclear reaction channel to provide summed nuclide and elemental distributions. It can also provide total (heavy) PKA distributions, which are particularly suitable for use as sampling distributions in atomistic modelling of radiation damage creation and evolution. Calculations of the summed PKA spectra as a function of material, neutron spectrum, and irradiation time reveal that there can be significant variation, which would, in turn, result in very different damage production rates.

For example, when elemental sum PKA distributions for Fe and W under a characteristic fusion reactor neutron spectrum are merged with damage functions -- as Frenkel-pairs (FPs) as a function of recoil energy -- calculated using the binary collision approximation (BCA), the results suggest that the rate of FP production is an order of magnitude higher in Fe. However, this approach, which uses the SRIM BCA code, is relatively simplistic as it does not consider evolution, and in this respect it is perhaps nearly equivalent to the damage quantifications based directly on the nuclear cross-section data, such as the displacements per atom (dpa) measure.
Hence we also consider another approach which samples the recoil energies from the global (heavy) sum PKA distributions according to the total PKA rate, inserts the appropriate distribution of defects, and simulates the evolution of objects via a stochastic cluster dynamics (SCD) model. The results for Fe and W under the same conditions suggest a completely opposite picture -- W defect production and clustering is much higher than in Fe. These results demonstrate the importance of considering evolution when attempting to quantify and compare radiation damage in materials because even using a relatively approximate defect evolution and clustering model produces a very different picture from that predicted by the BCA approach or similar.

\section{Acknowledgements}

M. Gilbert and J.-Ch. Sublet wish to thank Emmanuel Andrieu for exploring the SRIM code and testing the methodology by which results from SRIM could be combined with the present work. Thanks also to Sergei Dudarev for helpful discussions and suggestions. This project has received funding from the European Union's Horizon 2020 research and innovation programme under grant agreement number 633053 and from the RCUK Energy Programme [grant number EP/I501045]. J. Marian acknowledges support from the US-DOE's Early Career Research Program. To obtain further information on the data and models underlying this paper please contact PublicationsManager@ccfe.ac.uk. The views and opinions expressed herein do not necessarily reflect those of the European Commission.

%\section*{References}
%\bibliographystyle{mgilbert_plain_JNM}
\bibliographystyle{elsarticle-num}
\bibliography{recoil_spectra_paper}

\end{document}